\documentclass[a4paper,prd,amsfonts, amssymb, amsmath, reprint, showkeys, nofootinbib, twoside, superscriptaddress,11pt]{article}
\usepackage{jheppub}
\usepackage[english]{babel}
\usepackage[utf8]{inputenc}
\usepackage[colorinlistoftodos, color=green!40, prependcaption]{todonotes}
\usepackage{xcolor}
\usepackage{graphicx}

\usepackage{adjustbox, float}
\usepackage{placeins}
\usepackage[T1]{fontenc}
\usepackage{lipsum}
\usepackage{csquotes}
\usepackage{amsmath, amssymb, slashed, minitoc, amsthm,graphicx, caption, tensor, mathtools,wrapfig,bm,esint,subcaption,cancel,booktabs}
\begin{document}
\begin{flushright}
DCPT-20/17
\end{flushright}
\bibliographystyle{apsrev4-2}
\title{Novel Hairy Black Hole Solutions in Einstein-Maxwell-Gauss-Bonnet-Scalar Theory}

\author[a,b]{Callum L.\ Hunter}
\author[a]{Douglas J.\ Smith}
\emailAdd{c.l.hunter@soton.ac.uk}
\emailAdd{douglas.smith@durham.ac.uk}


\affiliation[a]{Department of Mathematical Sciences\\
Durham University, Upper Mountjoy, Stockton Road, Durham DH1 3LE, UK}
\affiliation[b]{Mathematical Sciences and STAG Research Centre, University of Southampton,\\
Highfield, Southampton, SO17 1BJ, UK}

\date{\today} 
\abstract{
It has been previously shown that a Gauss-Bonnet term non-minimally coupled to a scalar field produces a scalarised black hole solution which can be considered as having secondary scalar hair, parametrised in terms of the black hole's mass and charge. In this paper we extend a previously investigated linear coupling of the form $f(\phi)=\phi$ to a non-minimally coupled Maxwell term, with the form $\frac{1}{8}F_{\mu\nu}F^{\mu\nu}+\beta\phi F_{\mu\nu}F^{\mu\nu}$. By using numerical methods the solutions to the full differential equations are found, as well as a perturbative expansion in the $r\rightarrow\infty$ limit and a perturbative expansion in couplings parameters such as $\beta$. These solutions describe scalarised black holes with modified electric field which have dependence not only on the electric charge of the black hole, but also the value of the non-minimal coupling constant. We also discuss the bounds imposed on the parameters of the black hole by the reality condition of the solution, giving some explicit numerical bounds.}
\maketitle
\flushbottom
\section{Introduction}
With the advent of gravitational wave astronomy \cite{PhysRevLett.116.061102, PhysRevLett.116.241103, PhysRevLett.118.221101, PhysRevLett.119.141101, PhysRevLett.119.161101,Abbott_2017} there is a need to closely investigate our current theories of gravity in an astrophysical setting. One of the most fruitful areas to do this is the arena of black hole physics.

The hair, or rather lack thereof, of black holes has been well established in the literature assuming a General Relativistic-like theory of gravity \cite{PhysRev.164.1776, Israel1968, PhysRevLett.26.331, PhysRevD.5.1239, PhysRevLett.28.452}. In these theories a black hole can be characterised by only three Kerr-Newman quantities: the mass, electromagnetic charge and angular momentum. These \textit{no-hair} theorems have since been extended to Brans-Dicke and certain Scalar-Tensor theories \cite{PhysRevD.95.124013, PhysRevLett.108.081103}. In Brans-Dicke theory, it was shown that the end points of collapse to stationary black holes are identical to their GR counterparts \cite{Hawking1972}, which are described by the Kerr-Newman metric. These theorems have further been extended to some Galileon models of gravity \cite{PhysRevD.85.103501,PhysRevLett.110.241104}, where it was shown that using the `John' term \cite{PhysRevLett.108.051101} scalar hair could not be supported outside of the horizon\footnote{However, it is possible to support so-called `stealth hair' \cite{PhysRevD.100.084020}. This kind of hair leaves the metric solution unchanged when compared to GR, but a non-trivial scalar field is supported outside of the horizon. It is also possible to circumvent no-hair theorems if certain couplings of the scalar field are logarithmic or have negative powers of the scalar \cite{Babichev:2016rlq}.}. In this paper we are mainly concerned with the no-scalar hair arguments that are put forward in these previous works for spherically symmetric, static spacetimes.

The arguments for each of these no-scalar hair theorems follow the same logical structure as in \cite{PhysRevLett.110.241104}. That is, the scalar equation of motion can be written as a current conservation equation due to the shift symmetry implied by the Galileon model, hence $\nabla_\mu J^\mu=0$. Then, due to spherical symmetry and a static spacetime, the only non-zero component of the current is the radial part, of the generic form,
\begin{equation}\label{1}
    J^r=\frac{\phi'}{e^{B(r)}}\Gamma(\phi';g,g',g''),
\end{equation}
where $\Gamma$ is a generic function which depends upon the form of the action, $e^{B(r)}$ is the radial component of the spherically symmetric metric, $g$ denotes metric components and $'$ denotes the radial derivative. Then, using the conditions that the spacetime must be asymptotically flat to be well-defined and $J_\mu J^\mu$ must be finite at the horizon, it is shown that $J^r=0$ everywhere by using the Poincar\'e invariance of the theory. The arguments are also outlined in \cite{PhysRevD.90.124063BHhairexample}. 

There are of course exceptions to the rule, and these exceptions are by no means new. The Einstein-Yang-Mills action with an $SU(2)$ gauge group challenges the no-hair theorems \cite{PhysRevLett.61.141, PhysRevLett.64.2844} by showing that there are configurations of the matter fields which produce an asymptotically flat spacetime such that they are regular on the horizon of the black hole. It was also found that black holes can have Skyrmion hair \cite{LUCKOCK1986341, 2016NuPhB.913.1001D}, which manifests as a Skyrmion charge of the black hole. Furthermore, it was shown in \cite{PhysRevD.47.2242} that non-Abelian gauge theories can also generate hairy solutions. Specifically, it was shown that an $SU(2)$ gauge group coupled to a Higgs doublet and an Einstein--non-Abelian--Proca action would produce hair. Moreover, general Galileon models are significantly more complex than standard scalar-tensor theories, such as Brans-Dicke, and so they are quite frequently not entirely encompassed by the no-hair proofs.

Our main focus in this work, however, is the development of scalar hair in black hole solutions. Over the past decade a variety of works have been produced which show that scalar hair can exist within certain scalar-tensor theories. It was shown in \cite{PhysRevLett.120.131102, PhysRevLett.120.131103} that if a scalar field $\phi$ is non-minimally coupled to the Gauss-Bonnet invariant $\mathcal{G}$, then a nontrivial scalar field configuration could be supported outside the event horizon. It was further shown in \cite{PhysRevLett.121.101102,2019arXiv190508304A} that the same process occurs for non-minimal coupling to the Maxwell tensor. In both cases, the coupling was a general function $f(\phi)$, and the issues of stability were not addressed. In this context, stability relates to the tendency of black holes to collapse into a Kerr-Newman end state or some other novel end state. In certain theories, with $f(\phi)$ satisfying certain conditions, the end state of collapse can either be a Kerr-Newman like solution or a new solution. However, illustrative examples of the consequences of instability can be found in \cite{doi:10.1142/S0217732308028351} where it was shown that non-minimal coupling to Born-Infeld electrodynamics produced different black hole solutions depending upon the mass of the black hole. It is thermodynamically favourable, under certain conditions, to generate a scalarised solution to the field equations rather than a Schwarzschild-like solution. A study of the instability in Einstein-Gauss-Bonnet-scalar (EGBS) theory found very much the same results, that is in certain regions it is favourable to scalarise the solution as the Schwarzschild solution becomes unstable and the system bifurcates to give new scalarised solutions \cite{PhysRevD.98.024030}.

Aside from the interest in violations of the no-hair conjecture, scalar-tensor theories are also of cosmological interest. Brans-Dicke gravity has been shown to produce eras of inflation followed by eras of almost no growth, assuming a specific potential \cite{fujii2003scalar}. EGBS gravity has been used to investigate inflation, producing non-trivial results such as driving inflationary stages without the need of a slow roll approximation \cite{PhysRevD.98.083539}. Furthermore, Galileon gravity has been investigated extensively in cosmological settings \cite{PhysRevD.80.024037}, and was found to produce some results which fit better with cosmological data \cite{PhysRevD.90.023528} than standard GR alone. While these results do not directly relate to the scalar hair surrounding black holes, they motivate investigating any consequences of such theories, as any deviation from the expected Schwarzschild-like metrics could provide evidence for such deviations from GR.

To avoid discussion of stability here, we have chosen a linear coupling of $\phi$ to $\mathcal{G}$ and $F_{\mu\nu}F^{\mu\nu}$ as a linear scalar coupling to the Gauss-Bonnet term excludes Schwarzschild-like solutions \cite{PhysRevD.90.124063BHhairexample}, so such models do not suffer from instability arguments. The reason such a term generates scalar hair is due to the form of the non-minimally coupled term in the equations of motion. In this case, the current obeys the requirements for a usual black hole (asymptotically flat with finite current), but there is no trivial way to determine the asymptotic behaviour of the function $\Gamma$ in (\ref{1}). Since we want to avoid any Ostrogradsky instability, we can only have second derivatives in the equations of motion, furthermore in the $\phi\rightarrow constant$ limit we need the term to be divergence free so we recover GR. The only choice by Lovelock's theorem, is the Gauss-Bonnet invariant \cite{PhysRevLett.112.251102}.

The structure of this paper will closely follow that outlined in \cite{PhysRevD.90.124063BHhairexample, PhysRevLett.112.251102, BRIHAYE2019295}. In Section \ref{S2} we shall present the action, motivate its form and derive the equations of motion for a general metric. We then go on to specify a spherically symmetric ansatz and derive the component differential equations associated with each of these equations of motion. In Section \ref{S3} we investigate the asymptotic and near-horizon solutions, presenting the conditions the scalar field must satisfy to be regular at the horizon and for the solution to be asymptotically flat. In Section \ref{S4} we present the numerical solutions to the equations of motion and investigate their consequences. Finally, in Section \ref{S5} we draw our conclusions.

\section{The Action and Equations of Motion}\label{S2}

The action we shall investigate in this paper is an extension of the action in \cite{PhysRevD.90.124063BHhairexample, PhysRevD.98.104056}. We shall consider not only the Gauss-Bonnet term, but also the Maxwell term non-minimally coupled to the scalar field. This will allow us to generalise the theories previously studied and also understand the effects of varying parameters within the theory. The simplest model we can investigate, in parallel with \cite{PhysRevD.90.124063BHhairexample}, is
\begin{equation}\label{2}
    \begin{aligned}
    S=&\frac{M_P^2}{8\pi}\int d^4x \sqrt{-g}\bigg(\frac{R}{2}-\frac{1}{8}F_{\mu\nu}F^{\mu\nu}-\frac{1}{2}\partial_\mu\phi\partial^\mu\phi+\phi(\alpha\mathcal{G}-\beta F_{\mu\nu}F^{\mu\nu})\bigg),
    \end{aligned}
\end{equation}
as a result of the dimensions of the derivatives, the scalar field is dimensionless and the gauge field strength has dimensions of inverse length; hence $\alpha$ is a coupling constant of dimension length squared, and $\beta$ is a dimensionless coupling constant. In our work we shall use natural units such that $G_N=\frac{1}{4\pi\varepsilon_0}=e=c=1$, where $G_N$ is Newton's constant. This action can be connected to a heterotic string theory with couplings of the form $e^{\gamma\phi}$ \cite{PhysRevD.97.064032}, which is well known to produce scalar hair \cite{PhysRevD.54.5049Dilatonicstring1}. To make this connection, we may consider the theory to be an expansion in the coupling parameters $\alpha, \beta$. The existence of this action can also be argued from an Effective Field Theory (EFT) viewpoint; however we have not included all operators present at this order and hence we are working with some restricted EFT.

The equations of motion that result from this theory are relatively straightforward to derive, the only troublesome term being the Gauss-Bonnet term. Details of the variation of this term can be found in Appendix B of \cite{PhysRevD.90.124063BHhairexample}. The equations of motion resulting from (\ref{2}) are
\begin{align}
    \Box\phi+(\alpha\mathcal{G}-\beta F_{\mu\nu}F^{\mu\nu})&=0,\label{3}\\
    G_{\mu\nu}&=\Tilde{\mathcal{T}}_{\mu\nu},\label{4}\\
    \partial_\nu(\sqrt{-g}(\tfrac{1}{8}+\beta\phi) F^{\mu\nu})&=0,\label{5}
\end{align}
where (\ref{5}) is the same as eqn.(2.4) in \cite{2019arXiv190508304A} and $\Tilde{\mathcal{T}}_{\mu\nu}$ represents the modified stress-energy tensor, and is given by,
\begin{equation}
    \begin{aligned}
    \Tilde{\mathcal{T}}_{\mu\nu}=&\partial_\mu\phi\partial_\nu\phi-\frac{1}{2}g_{\mu\nu}(\partial\phi)^2-\alpha(g_{\rho\mu}g_{\delta\nu}+g_{\rho\nu}g_{\delta\mu})\epsilon^{\lambda\eta\rho\sigma}\epsilon^{\alpha\beta\gamma\delta}R_{\lambda\eta\alpha\beta}\nabla_\sigma(\partial_\gamma\phi)\\
    &+(\tfrac{1}{2}+4\beta\phi)(F_{\mu\beta}\tensor{F}{_\nu^\beta})-g_{\mu\nu}(\tfrac{1}{8}+\beta\phi) F_{\alpha\beta}F^{\alpha\beta}.
    \end{aligned}
\end{equation}
The $\epsilon^{\lambda\eta\rho\sigma}\epsilon^{\alpha\beta\gamma\delta}R_{\lambda\eta\alpha\beta}$ term is the double dual Riemann tensor \cite{misner2017gravitation}, and is divergence free.
 Using the equations of motion (\ref{3}-\ref{5}) we can derive the component equations of motion in a similar way to \cite{PhysRevD.90.124063BHhairexample,PhysRevLett.120.131102}, and find the form of the electric field. By considering a spherically symmetric ansatz for the metric of the form,
 \begin{equation}\label{7}
    \begin{aligned}
     ds^2=-e^{A(r)}&\textnormal{d}t^2+e^{B(r)}\textnormal{d}r^2+r^2(\textnormal{d}\theta^2+\sin^2\theta\textnormal{d}\varphi^2),
     \end{aligned}
 \end{equation}
we obtain a time-independent, spherically symmetric scalar field of the form,
\begin{equation}\label{8}
    \phi=\phi(r).
\end{equation}
Next, we can reasonably make the ansatz that, allowing for the existence of electric monopoles only, the gauge potential has the form,
\begin{equation}\label{9}
    \mathcal{A}=V(r)\textnormal{d}t,
\end{equation}
where $V(r)$ is the electric potential and we shall denote the electric field by $E(r)$. As a result, the only nonzero components of the field strength tensor are $F_{01} = -F_{10} = E(r)$. We note briefly that our metric ansatz (\ref{7}) is only valid outside the horizon and to extend beyond the horizon we would have to multiply the $\textnormal{d}t^2$ components by -1.  We also note that the scalar field equation is redundant as it can be retrieved from the divergence-free nature of the modified stress energy tensor, $\nabla^\mu\tilde{\mathcal{T}}_{\mu\nu}=0$, as in \cite{PhysRevD.90.124063BHhairexample}. However, the scalar equation will be used extensively in the $r\rightarrow\infty$ expansion in Section \ref{S3}, and in the numerical solutions, as it is a simpler equation which can be used more conveniently in \textit{Mathematica}.
\subsection{The Scalar Equation}
We begin by finding the scalar equation of motion. The Gauss-Bonnet invariant, which appears in (\ref{3}) is given by the expression,
\begin{equation}\label{10}
    \begin{aligned}
    \mathcal{G}&=\frac{2(1-e^B)A'^2+2(e^B-3)A'B'}{r^2e^{2B}}+\frac{4(1-e^B)A''}{r^2e^{2B}},
    \end{aligned}
\end{equation}
and the d'Alembert operator acting on the scalar field gives,
\begin{equation}\label{11}
    \begin{aligned}
    \Box\phi&=\frac{1}{\sqrt{-g}}(\partial_\mu\sqrt{-g}g^{\mu\nu}\partial_\nu)\phi\\
    &=\frac{2\phi'}{re^B}+\frac{(A'-B')\phi'}{2e^B}+\frac{\phi''}{e^B}.
    \end{aligned}
\end{equation}
Finally, the field strength tensor squared is given by,
\begin{equation}\label{12}
    F_{\mu\nu}F^{\mu\nu}=-2E(r)^2e^{-A-B}.
\end{equation}
Combining (\ref{10}-\ref{12}), we obtain a modified version of the scalar field equation in \cite{PhysRevD.90.124063BHhairexample},
\begin{equation}\label{13}
    \begin{aligned}
    0&=8\alpha(e^{-B}-1)A''+2r^2\phi''+r^2(A'-B')\phi'+4\alpha(1-3e^{-B})A'B'\\
    &\qquad+4\alpha(e^{-B}-1)A'^2+4r\phi'+4\beta r^2e^{-A}E(r)^2,
    \end{aligned}
\end{equation}
which in the limit of $F^2\rightarrow0$ recovers the exact equation found in \cite{PhysRevD.90.124063BHhairexample}.
\subsection{The Electric Field Equation}
Using the ansatz of (\ref{9}) in (\ref{5}), and noting that $V'(r)=E(r)$, we can determine the equation governing the electric field in terms of the radial coordinate, scalar field and metric components. Identifying the integration constant resulting from integrating (\ref{5}) as the electric charge of the black hole, we obtain the solution,
\begin{equation}\label{14}
    E(r)=\frac{e^{\frac{A+B}{2}}}{r^2}\frac{Q}{(1+8\beta\phi)},
\end{equation}
which is only singular at $r=0$, provided that $\phi\neq-\frac{1}{8 \beta}$.
In fact we must take the bound $\beta \phi>-\frac{1}{8}$ as we can see from the action (\ref{2}) in order for the energy to be bounded from below. In particular, as $\phi$ cannot cross the value where $\beta\phi = -\frac{1}{8}$, $\beta\phi$ must everywhere be either above or below $-\frac{1}{8}$, but in the case that it is below the contribution from the gauge field has the wrong sign. Indeed, since we necessarily must impose asymptotic boundary conditions $\phi \rightarrow 0$ we again see that we must have $\beta\phi > -\frac{1}{8}$. This suggests that $\beta\phi=-\frac{1}{8}$ is a bifurcation point of solutions in the theory leading to two branches, those with $\beta\phi>-\frac{1}{8}$ and those with $\beta\phi<-\frac{1}{8}$ which are nonphysical. Note that if $\beta > 0$ we then have $\phi > -\frac{1}{8\beta}$ while instead if $\beta < 0$ we then have $\phi < -\frac{1}{8\beta}$. It turns out that finding solutions in the case where $\beta<0$ is numerically difficult due to divergences for some initial conditions which cause issues in the automatic shooting method used in \textit{Mathematica}. However we were able to find some approximate solutions, presented in Section \ref{S4}, in this regime by using a manual shooting method and searching for solutions by hand.

The key thing to note here is that this equation highlights the explicit interdependence of the electric and scalar fields and demonstrates that the scalar field serves to suppress the electric field outside the horizon. The suppression will become obvious in the numerical results presented in Section \ref{S4} (see Figure \ref{fig6}) and in the asymptotic expansion of the fields, as the electric field has a correction of order $r^{-3}$ which is proportional to the scalar charge.

\subsection{The Components of the Einstein Equation}
Next we shall find the $tt,rr$ and $\theta\theta$ components of the Einstein equation (\ref{4}). We shall explicitly show the form of the stress energy tensor and Einstein tensor for the $tt$ case, and then present the rest of the equations.

We begin with the Einstein tensor for the $tt$ component, which is the same as in \cite{PhysRevD.90.124063BHhairexample},
\begin{equation}\label{15}
    G_{tt}=\frac{e^{-A}}{r^2}(e^{-B}B'r-e^{-B}+1).
\end{equation}
The stress energy tensor is more complicated as it contains the double dual Riemann tensor and the second-order derivatives of the scalar field. When the smoke clears, and we simplify as far as we can, the stress energy tensor component takes on the form,
\begin{equation}\label{16}
    \begin{aligned}
    e^{-A}\tilde{\mathcal{T}}_{tt}&=\frac{4\alpha(e^{B}-3)\phi'B'}{r^2e^{2B}}-\frac{\phi'^2}{2e^B}-\frac{8\alpha(e^B-1)\phi''}{r^2e^{2B}}-(\tfrac{1}{8}+\beta\phi) E(r)^2e^{-A-B}.
    \end{aligned}
\end{equation}
This leads to the $tt$ equation of motion,
\begin{equation}\label{17}
    \begin{aligned}
    0&=16\alpha(1-e^{-B})\phi''+8\alpha(3e^{-B}-1)\phi'B'+r^2\phi'^2+         4(\tfrac{1}{8}+\beta\phi)e^{-A}E(r)^2r^2\\
    &\quad-2rB'-2e^B+2,
    \end{aligned}
\end{equation}
which is a nonlinear coupled equation for the metric and scalar field. This is a general theme of this type of gravitational theory: non-linear coupled differential equations which can only be solved numerically or perturbatively.
The $rr$ component then follows the same sort of analysis, and gives the equation,
\begin{equation}\label{18}
    \begin{aligned}
    0&=(e^B)^2+e^B\big[\tfrac{\phi'^2r^2}{2}-4\alpha\phi'A'-(A'r+1)-2(\tfrac{1}{8}+\beta\phi)E(r)^2r^2e^{-A}\big]+12\alpha\phi'A',
    \end{aligned}
\end{equation}
which is a quadratic in the metric function $e^B$. This fact was used in \cite{PhysRevD.90.124063BHhairexample,PhysRevLett.120.131102} in order to derive two coupled differential equations for $A''$ and $\phi''$, which is the method we adopt in this case, and can be found in Section \ref{S3}.

Finally, we have the $\theta\theta$ equation, which is the same as the $\varphi\varphi$ equation due to the assumed spherical symmetry of the solution,
\begin{equation}\label{19}
    \begin{aligned}
    0&=16\alpha(\phi'A''+\phi''A')r-2A''r^2e^{B}+8\alpha(A'-3B')A'\phi'r-2\phi'^2r^2e^B\\
    &\quad-A'(A'-B')r^2e^B-2(A'-B')re^B+(1+8\beta\phi)r^2E(r)^2e^{B-A}.
    \end{aligned}
\end{equation}
As a basic check, we note that all the equations of motion become the expected equations in \cite{PhysRevD.90.124063BHhairexample} in the limit of smoothly sending $\beta$ and $Q$ to zero.

Next we algebraically solve the $rr$ equation for $e^B$ to find
\begin{equation}\label{20}
    e^B=\frac{-\Lambda\pm\sqrt{\Lambda^2-4\Delta}}{2},
\end{equation}
where we have defined,
\begin{equation}\label{21}
    \begin{aligned}
    \Lambda&=\big[\tfrac{\phi'^2r^2}{2}-4\alpha\phi'A'-(A'r+1)-2(\tfrac{1}{8}+\beta\phi)E(r)^2r^2e^{-A}\big],
    \end{aligned}
\end{equation}
and,
\begin{equation}\label{22}
    \Delta=12\alpha\phi'A'.
\end{equation}
We take the positive root of this equation as this ensures the correct asymptotic behaviour of the $e^{B(r)}$ solution. This equation can then be used to eliminate $B'$ and $e^B$ from the other equations of motion, which subsequently can be solved numerically, as we will do in Section \ref{S4}. The electric field $E(r)$ depends upon $e^B$, and hence when we find numerical solutions we shall substitute \eqref{14} in \eqref{19} and then solve for $e^B$ again. Alternatively, we could first substitute \eqref{14} into \eqref{19} to eliminate $E$ and then solve for $e^B$.

\section{The Limiting Cases}\label{S3}
In this section, we investigate the behaviour of the solution at the horizon and at infinity. The former limit will give us a boundary condition for the gradient of $\phi'$ at the horizon, and the latter limit will give us an instinctive feeling for the behaviour of the solutions as seen by an asymptotic observer, and will aid us in Section \ref{S4} in the case when $\beta<0$. We ultimately expect the asymptotic solution to produce something close to the Reissner-Nordstr\"om solution to order $\mathcal{O}(\frac{1}{r^2})$, since in the small $\phi$ limit of the action (\ref{2}) we approximately get the Einstein-Maxwell action. We shall use this fact in Section \ref{3B} in order to fix some of the parameters within the infinite perturbative expansion.

\subsection{The Near Horizon Limit}
To find black hole solutions, we assume there is a horizon at $r=r_h$. At the horizon, the metric must satisfy certain properties \cite{PhysRevLett.120.131102}, including that as $r\rightarrow r_h$ we must have $e^A\rightarrow0$ which implies that $e^{-A}\rightarrow\infty$ and $A'\rightarrow\infty$. In performing this analysis, there are two approaches. The first is to analyse the behaviour of $e^B$ near the horizon and then use this to determine the near horizon behaviour of $A''$ and $\phi''$ \cite{PhysRevD.90.124063BHhairexample, PhysRevLett.120.131102, PhysRevD.97.084037}. The other option is to use an explicit expansion of the metric function as in \cite{PhysRevD.54.5049Dilatonicstring1, PhysRevD.57.6255Dilatonicstring2}. While the latter method gives an explicit expansion of the metric functions near the horizon, it is rather complicated. Hence here we shall analyse the behaviour of $e^B$ and use this to determine the near horizon behaviour. In order to do this, we shall investigate the behaviour of (\ref{20}) in the near horizon regime, finding the leading order term in an expansion of the square root. We first begin by noting that in this limit $e^{-A}\approx A'/a_1$ due to the metric function form near the horizon as shown in more detail in Appendix \ref{AA}. Using this fact, and expanding to $\mathcal{O}(1)$, we can see that (\ref{20}) takes on the approximate form,
\begin{equation}\label{23}
    \begin{aligned}
    e^B=(&r+4\alpha\phi'+\tfrac{2}{a_1}(\tfrac{1}{8}+\beta\phi)E(r)^2r^2)A'-\bigg(\frac{1}{2}\phi'^2r^2\\
    &+\frac{8\alpha\phi'-r-2(\tfrac{1}{8}+\beta\phi)(a_1)^{-1}E(r)^2r^2}{r+4\alpha\phi'+2(\tfrac{1}{8}+\beta\phi)(a_1)^{-1}E(r)^2r^2}\bigg)+\mathcal{O}\bigg(\frac{1}{A'}\bigg),
    \end{aligned}
\end{equation}
which replicates eqn.(36) in \cite{PhysRevD.90.124063BHhairexample} in the $E(r)\rightarrow0$ limit. Next, we substitute this into (\ref{13}), (\ref{17}) and (\ref{19}) to produce three simultaneous equations, two of which can be used to produce the near-horizon $A''$ and $\phi''$ equations mentioned earlier, with the third serving as a consistency check. Given that in this process we have eliminated all divergent factors in favour of terms proportional to $A'$, we can take an asymptotic expansion of these equations about $A'\rightarrow\infty$ which yields,
\begin{align}
    \phi''&=f(\phi, \phi', A; \alpha, \beta)A'+\mathcal{O}(1)\label{24},\\
    A''&=g(\phi, \phi', A; \alpha,\beta)A'^2+\mathcal{O}(A')\label{25},
\end{align}
where the function $g$ is not important to our current analysis as we do not explicitly consider the behaviour of $A''$ near the horizon. We do, however, require the form of the function $f$ as this will determine our initial gradient for the numerical solutions in Section \ref{S4}. The function $f$ for $\phi''$ is given by (\ref{B4}) in Appendix \ref{AB}, and whilst this equation is tedious it does reproduce the expected result from \cite{PhysRevD.90.124063BHhairexample, PhysRevD.99.064003} and hence serves as an important check for our work. In order for $\phi''$ to remain finite on the horizon, we must have that the second bracket in the numerator of (\ref{B4}) is identically zero. We can then solve that equation for $\phi'$, in doing so we obtain,

\begin{equation}
    \begin{split}
    \phi'_h=&-\Big[-64 \alpha  \beta  Q^2 r_h^4+8 \alpha ^2 \big(4 Q^2 r_h^2 (8 \beta  \phi _h+1)+Q^4\big)+\big(Q^2-4 r_h^2 (8 \beta  \phi _h+1)\big)\\
    &\times\Big(r_h^6 (8 \beta  \phi _h+1)\mp\big\{64 \alpha ^4 Q^2 \big(24 r_h^2 (8 \beta  \phi _h+1)+Q^2\big)-1536 \alpha ^3 \beta  Q^2 r_h^4\\
    &+32 \alpha ^2 r_h^6 (8 \beta  \phi _h+1) \big(Q^2-6\alpha^2 r_h^2 (8 \beta  \phi _h+1)\big)+r_h^{12} (8 \beta  \phi _h+1){}^2\big\}^{\frac{1}{2}}\Big)\Big]\\
    &\Big/\Big[8 \alpha  \Big(32 \alpha ^2 Q^2 (8 \beta  \phi _h+1)-32 \alpha  \beta  Q^2 r_h^2+r_h^4 (8 \beta  \phi _h+1) \big(Q^2-4 r_h^2 (8 \beta  \phi _h+1)\big)\Big)\Big],\label{26}
    \end{split}
\end{equation}
which we shall use as an initial condition with a free value for $\phi$ at the horizon. We take the negative sign of the square root as this is required for a smooth $\alpha \rightarrow 0$ limit, noting that the denominator vanishes in this limit. The required value of $\phi$ to solve the initial condition will then be determined by the shooting method.

In order to deal with the presence of the electric field in the initial condition, we substitute in (\ref{14}) and then specify the electric charge for each of the black hole solutions found. The result in (\ref{26}) is very similar in structure to Eqn.(33) in \cite{PhysRevD.98.104056} which gives a scalarised Reissner-Nordstr\"om black hole; the action is similar to that in (\ref{2}), however there is no non-minimal coupling between $\phi$ and the gauge field. The complexity of (\ref{26}) is one of the reasons using the shooting method in this theory is difficult, the initial gradient is very sensitive to small changes in the initial values of the parameters. Now that we have analysed the near horizon behaviour, we can move on to investigating the approximate behaviour of the solution at spatial infinity. This will demonstrate the behaviour at very large $r$.

\subsection{Perturbative Solutions at Infinity}\label{3B}
In this subsection, we analyse the asymptotic solutions via a perturbation method. To apply this, first we must lay out what we expect to happen at infinity. The spacetime must be asymptotically flat to be a stable solution to Einstein's equations, and so we enforce that as $r\rightarrow\infty$ the metric tends to the Minkowski metric, $g_{\mu\nu}\rightarrow\eta_{\mu\nu}$. Hence, in this limit, $e^A\rightarrow1$, $e^B\rightarrow1$ and the scalar field must become a constant, i.e.\ $\phi' \rightarrow 0$. We can also set the asymptotic value of $\phi$,  $\phi(\infty)=0$. These assumptions are also outlined in \cite{PhysRevLett.120.131102, PhysRevD.97.084037, PhysRevD.99.064003}. We may then also use a power series expansion in $1/r$ in order to expand at spatial infinity,
\begin{align}
    e^{A(r)}&=1+\sum_{n=1}^\infty\frac{\tilde{a}_n}{r^n},\label{27}\\
    e^{B(r)}&=1+\sum_{n=1}^\infty\frac{\tilde{b}_n}{r^n},\label{28}\\
    \phi(r)&=\sum_{n=1}^\infty\frac{\Tilde{\phi}_n}{r^n},\label{29}\\
    E(r)&=\sum_{n=2}^\infty\frac{\tilde{q}_n}{r^n},\label{30},
\end{align}
where we shall determine the constants order-by-order. In order to do this we first fix three of the constants, since we have a reasonable idea of the behaviour of the expansion of $e^A$, $\phi$ and $E(r)$. Specifically, $\tilde{a}_1=-2M$, we can also fix $\Tilde{\phi}_1=P$ where $P$ is the scalar charge of the black hole, and finally $\tilde{q}_2=Q$ with $Q$ being related to the electric charge of the black hole. These choices are based upon the assumption that in a $1/r$ expansion, to leading order the solutions will match those of the Reissner-Nordstr\"om solution.

We can then substitute (\ref{27}-\ref{30}) into the equations of motion (\ref{13},\ref{17}-\ref{19}), and solve the equations order by order using \textit{Mathematica}, finding
\begin{align}
    e^{A(r)}\approx&1-\frac{2M}{r}+\frac{Q^2}{4r^2}+\frac{MP^2-4\beta PQ^2}{6r^3} \label{31}\\
    e^{B(r)}\approx&1+\frac{2M}{r}+\frac{16M^2-2P^2-Q^2}{4r^2}+\frac{16M^3-5MP^2-2MQ^2+4\beta PQ^2}{2r^3}, \label{32}\\
    E(r)\approx&\frac{Q}{r^2}-\frac{8\beta PQ}{r^3}+\frac{-32\beta MPQ-P^2Q+256\beta^2P^2Q+32\beta^2Q^3}{4r^4} \label{33}\\
    \phi(r)\approx&\frac{P}{r}+\frac{MP-\beta Q^2}{r^2}+\frac{16M^2P-P^3-PQ^2+64\beta^2 PQ^2-16\beta MQ^2}{12r^3}\label{34}
\end{align}
where all symbols have the usual meaning. Here we have displayed terms up to order $1/r^3$ in the metric components and $1/r^4$ in the electric field. The next order terms can be found in Appendix \ref{AB}, since after this order the numerator expressions become unwieldy and long. Turning off the electric field, (\ref{28}-\ref{31}) become eqns.(31-33) in \cite{PhysRevD.90.124063BHhairexample} as expected. We have three constants to specify for a physical black hole: $M$, $P$ and $Q$. Hence we would expect the family of solutions to be specified by the triplet $(M,P,Q)$, so three initial conditions are needed. This is not unexpected, as explained in \cite{PhysRevD.90.124063BHhairexample}, since for a given $r$ we must specify $A'$, $\phi'$ and something about the electric field $E$. We also recover the expected asymptotic behaviour for a black hole in a flat spacetime.

\subsection{Expanding in \texorpdfstring{$\alpha$}{Lg}}\label{alphaexpansion}
Performing an expansion in small $\alpha$ as in \cite{PhysRevD.83.104002Nonspinning} is not possible in this scenario as we do not have a zeroth order analytic solution to begin the expansion.
That is, assuming the zeroth order metric is the Reissner-Nordstr\"om solution, there are no closed forms for $e^A, e^B$ or $\phi$ with the electric field taking the form in (\ref{14}). That being said, it was possible to extract some information about the $\alpha$ expansion by considering expansions about the Schwarzschild metric, and hence treating the $Q$ value as a small perturbation from the Schwarzschild solution. In doing so, we can find the $\mathcal{O}(\alpha)$ correction to the scalar field $\phi$. If we then match this with the perturbative solutions at infinity, we find that,
\begin{equation}\label{35a}
    P\approx\frac{2\alpha+8\beta\phi_hQ^2}{M},
\end{equation}
which agrees with \cite{PhysRevD.90.124063BHhairexample} in the limit $Q\rightarrow0$.
This is only an approximation for small $Q^2$, $\alpha$ and $\beta$. However, it does give some explanation of the numerical results in the following section. These numerical results go some way to confirm that an expansion from Schwarzschild spacetime, for $Q=\mathcal{O}(0.1)$, does give an answer that provides the correct scaling laws for the value of the scalar charge.

In the situation where $Q\neq0$, the leading term of the scalar charge has a dependence on the scalar field's horizon value $\phi_h$. This suggests that the scalar hair in this system is not necessarily completely secondary, that is the scalar charge cannot be expressed in terms of the Kerr-Newman quantities only. In order to express the scalar charge, a property of the scalar field solution, we need to know something about the scalar field solution, thus the hair may not necessarily be secondary. The $xTras$ package \cite{NUTMA20141719} for \textit{Mathematica} was used in deriving (\ref{35a}).

\section{Numerical Solutions}\label{S4}
In this section we present the numerical solutions to the differential equations given in Section \ref{S2}. Note that in the Subsections \ref{SSB} and \ref{SSC} we take $\alpha=\beta$ in order to explore the behaviour of varying charge, mass and overall coupling. We note that since $\alpha$ and $\beta$ have different dimension there is a dimensional constant that relates the two. Since the only dimensional quantity in \eqref{2} is $M_P^2$, the dimensionally correct relation between $\alpha$ and $\beta$ will be proportional to $M_P^2$. In subsection \ref{SSE} we take $\alpha\neq\beta$ in order to explore the effects of different couplings between the scalar and gravity, and the scalar and the electric field. In subsection \ref{negb} we investigate the case where we take $\beta<0$. Throughout this section we set $r_h=1$ since the solutions are given in the dimensionless parameter $r/r_h$ and $r_h$ can be arbitrarily chosen, along with the other parameters, to give a black hole with the desired mass.

\subsection{The Method}\label{SSA}

In order to solve the equations numerically, we implemented \textit{Mathematica}'s shooting method function within \textit{NDSolve}. The implementation required using the gradient (\ref{26}), however we may express this in a clearer way.  The form of this gradient is given by,
\begin{equation}
\phi'_h=\frac{\psi+\varphi\sqrt{\chi}}{\zeta},
\label{phiprime_h}
\end{equation}
where we have given $\psi,\varphi,\zeta$ in Appendix \ref{AB}, and can be determined from \eqref{26}. However, $\chi$ imposes a reality condition on the gradient of the scalar field, and we can use this to check if our parameters give rise to a solution. The form of $\chi$ is,
\begin{equation}
\begin{aligned}
\chi = &
r_h^8(r_h^4 - 192\alpha^2) + 32 \alpha^2 Q^2 r_h^6 + 1536\alpha^4 Q^2 r_h^2 (1 + 8\beta \phi_h)^{-1} \\
 & + 64\alpha^3 Q^2 (\alpha Q^2 - 24\beta r_h^4)(1 + 8\beta \phi_h)^{-2}
\geq 0,
    \end{aligned}
\label{chi_phiprime_h}
\end{equation}
which contains the value of the scalar field at the horizon $\phi_h$. In the $Q\rightarrow0$ limit we recover the results of \cite{PhysRevD.90.124063BHhairexample} for the function of $\phi'_h$.
The equation in \eqref{chi_phiprime_h} is order 12 in $r_h$ and order $5$ in $\alpha=\beta$, thus it cannot be analytically solved for $r_h$ or $\alpha=\beta$ using \textit{Mathematica}. We can solve for $\alpha$ and $\beta$ when they are not equal but the equations do not give us any more information, the same is true for solving for $\phi_h$. However, the value of this scalar field will be $\mathcal{O}(1)$ and so we shall take $\phi_h\approx1$\footnote{Numerical tests in \textit{Mathematica} show the explicit value of $\phi_h$ does not affect the limits of the inequalities of the parameters a great deal. Corrections are of order $0.001$ for $\alpha$ and $0.01$ for $r_h$, while there is no effect on $Q$ in the range of $\alpha$, $\beta$ and $r_h$ we shall be testing as it always satisfies the inequality \eqref{chi_phiprime_h}}. Under this assumption we can solve for $Q$, and in doing so we find that $Q\in\mathbb{R}$ for all acceptable values of $\alpha$, $\beta$ and $r_h$.

With this new $\phi'_h$, we can then express the equations of motion in terms of a tortoise coordinate of the form,
\begin{equation}\label{37}
    x=1-\frac{r_h}{r},
\end{equation}
as this allows us to investigate the full range of the solution by looking at the open interval $x=(0,1)$. The interval for $x$ cannot be closed as this leads to issues near the horizon and at infinity due to the numerical integration technique. We found that we can consistently solve the equation of motion on the interval $x=[0.01,0.999]$, which gives a numerical range of $r=[1.01,1000]r_h$. While this is not the full range, most of the interesting effects are displayed within $\sim10r_h$ of the singularity.

In order for the shooting method to produce a solution, we first implemented \textit{NDSolve} using an approximation to the initial gradient and values of the functions. The data was then extracted and used in the shooting method implementation. We repeated this process for each set of parameters we took within the parameter space. There are numerical errors associated with the shooting method employed here. To examine these errors, we plotted the differential equations derived above and looked at regions where these equations were not satisfied by the solutions found, see Appendix \ref{AppC}. The example in the appendix shows that the residuals errors of these differential equations were around 0.2\%, hence the numerical results in this paper can be trusted to a high degree of accuracy.

The parameters we chose to use were $(\alpha, \beta,Q)$ as they characterised the solution's non-minimal coupling strength and electric charge. Hence, we investigate the variations of each of the parameters with the other values held fixed in order to extract some qualitative information about the solution.

\subsection{Varying \texorpdfstring{$Q$}{Lg}}\label{SSB}

\begin{figure}

\begin{subfigure}{0.49\textwidth}
    \captionsetup{width=\textwidth}
    \centering
    \includegraphics[width=\textwidth]{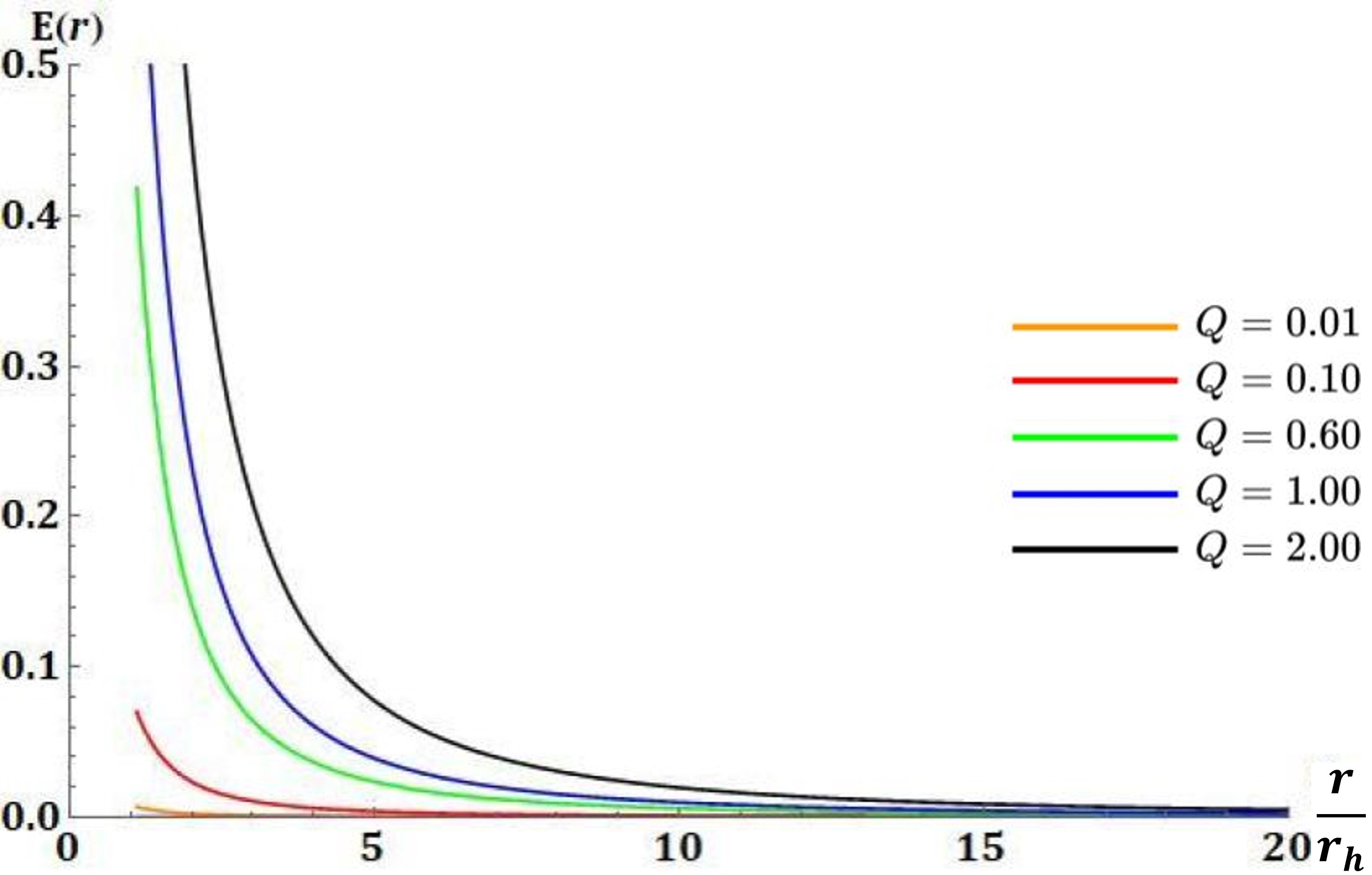}
\caption{The Electric field for various values of $Q$. Unsurprisingly the value of $Q$ has a major effect on the Electric field.}\label{fig1a}
\end{subfigure}
\hfill
\begin{subfigure}{0.49\textwidth}
    \captionsetup{width=\textwidth}
    \centering
    \includegraphics[width=\textwidth]{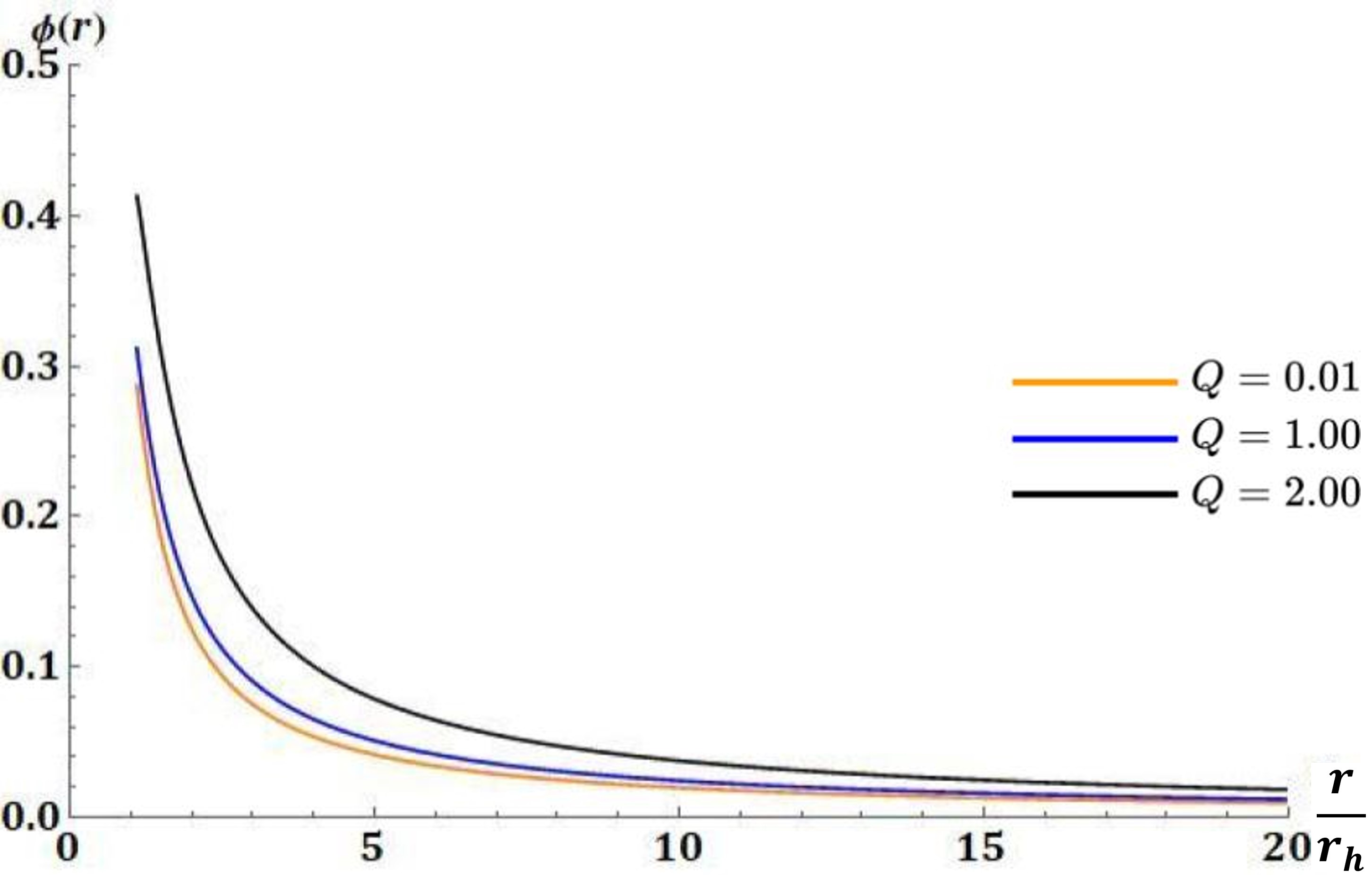}
\caption{The scalar field for various values of $Q$, once $Q>1$ the changes in charge have larger effects as expected from (\ref{35a}).}\label{fig1b}
\end{subfigure}
\caption{The field results of varying $Q$ with fixed $\alpha=\beta=0.05$ and $r_h=1$.}\label{fig1}

\end{figure}

In this subsection, we present the results for various values of $Q$. One issue we should first tackle regarding our results is the problem of `overcharging', which occurs in the Reissner-Nordstr\"om metric for $2Q_{RN}>r_s$. Using our third order perturbative expansion, (\ref{31}), set to 0, we approximated the bound and checked it numerically for each of the values given in the solutions. It was found that all solutions, in this subsection and the consequent subsections, had real solutions\footnote{We took an imaginary part of order $10^{-12}$ or less to be 0 due to numerical errors and the approximation of the bound.}.

We fix the values $\alpha,\beta=0.05$ throughout the variation of $Q$ so we can investigate the results of charge variation only. Using (\ref{chi_phiprime_h}) we find that all real values of $Q$ are acceptable for this $\alpha$ and $r_h$, as expected. The only thing we are limited by is our computing power, and so we only investigate the range $0.01\leq Q\leq2$. 

\begin{figure}
    \captionsetup{width=0.9\linewidth}
    \centering
    \includegraphics[width=0.6\linewidth]{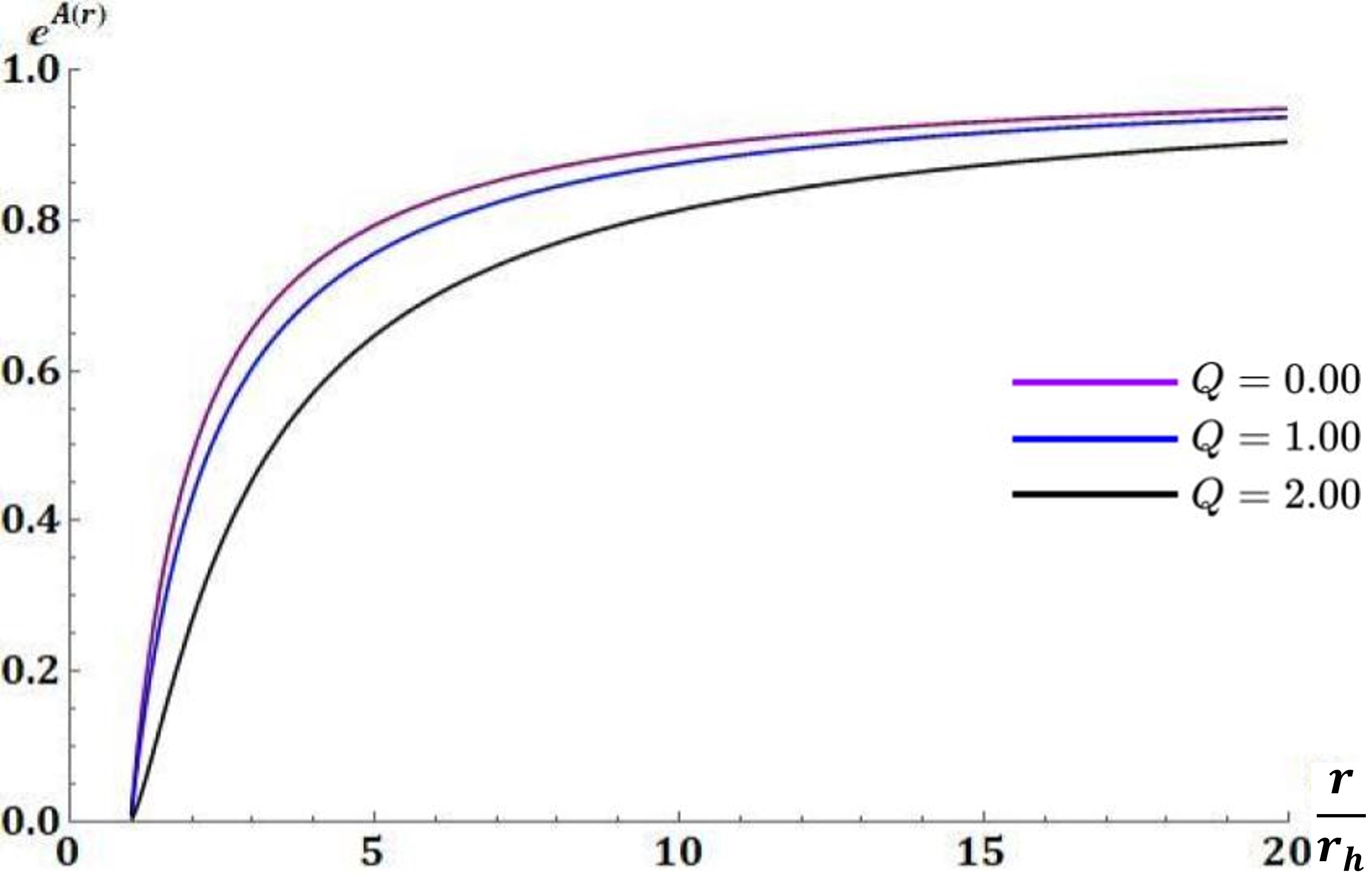}
\caption{The time component of $ds^2$ for various values of $Q$. Once $Q>1$ the changes in charge have larger effects.}\label{fig2}
\end{figure}

The results of the $Q$ variation are displayed in Figures \ref{fig1} and \ref{fig2}. The main effect we can see is with the $E(r)$ field, which is not surprising since this field is mainly sourced via the charge of the black hole. We also note that, as can be seen explicitly in Figure \ref{fig1a}, the Electric field is finite at the value $r/r_h=1$ and so the asymptotic conditions which were imposed in the no-hair theorem are satisfied here. The scalar field results are given in Figure \ref{fig1b}. Again the asymptotic conditions we require for a black hole solution are satisfied, that is, the field is finite across the horizon and as $r\rightarrow\infty$, $\phi \rightarrow 0$. It is interesting to note that the charge of the black hole does not have a considerable effect on the value of the scalar field. However, when $Q>1$ the effects become amplified. For values $0.1<Q<1$, not shown here, the effects are not as great as $Q>1$, and further analysis shows that for ever smaller values of $Q$ we asymptotically approach the $Q=0$ value. This is somewhat expected due to the approximate form of the scalar charge given in $(\ref{35a})$, where $P\propto Q^2$ and so it is not wholly unexpected that for $Q>1$ the scalar field increases in strength faster than the $Q<1$ case. Finally, the same effects are observed in the time component of the metric, that is, we have asymptotically flat solutions and the effects of varying $Q$ only become large for $Q>1$.
\subsection{Varying \texorpdfstring{$\alpha=\beta$}{Lg}}\label{SSC}
\begin{figure}
    \centering
    \captionsetup{width=0.9\linewidth}
    \centering
    \includegraphics[width=0.6\linewidth]{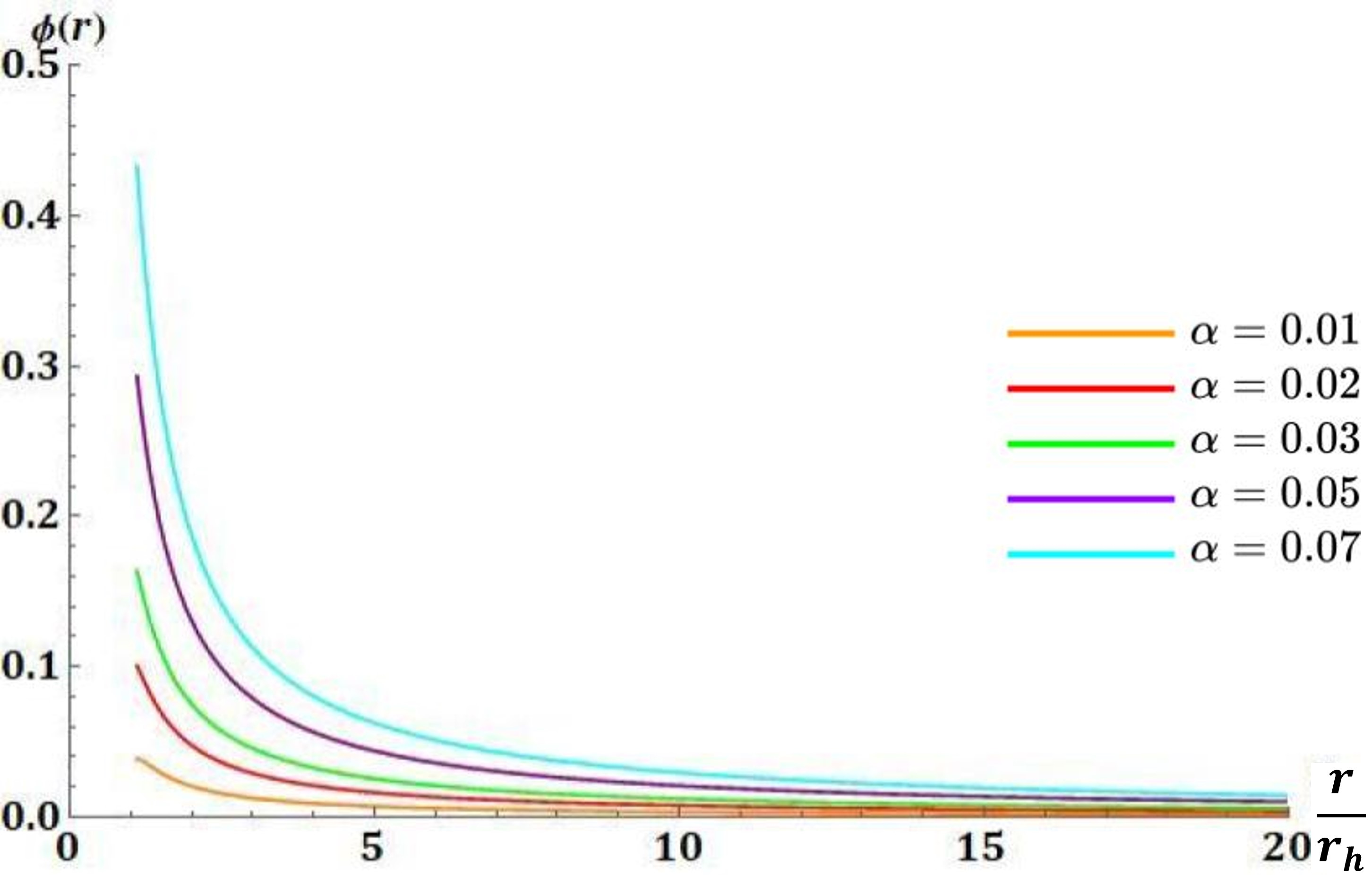}
    \caption{The $\phi$ field plotted for various values $\alpha=\beta$. As expected, larger values of $\alpha=\beta$ produce a larger scalar field.}
    \label{fig3}
\end{figure}

In varying $\alpha=\beta$ we kept $Q=0.5$; we also restrict $\alpha \lesssim 0.07$ as this is the approximate limit on $\alpha$ so that the bound in equation~(\ref{chi_phiprime_h}) is satisfied, as required for the gradient of $\phi$ to be real at the horizon as given by equation~(\ref{phiprime_h}). The results of these variations can be seen in Figure \ref{fig3}. We first note that varying the coupling for this $(\alpha,Q)$ does not produce a great deal of variation with respect to the electric field and the time component of the metric.

Furthermore, we only focus on values between $0.01\leq\alpha\leq0.07$ as values below $0.01$ produce only very small scalar fields and do not add much to the discussion. In the case of $\alpha\sim\mathcal{O}(0.001)$ the scalar profile does not change for varying values of $\alpha$ and so we can only conclude that the numerical errors following from the solution are larger than the contribution from the actual scalar field.
It seems that the increase in the scalar field is linear with the increase in the coupling constant $\alpha$. This follows from the form of $(\ref{3})$ and $(\ref{35a})$, which are linear in $\alpha$ and hence it is not unexpected that the $\phi$ field increases linearly with the value of $\alpha$.

\subsection{Taking \texorpdfstring{$\alpha\neq\beta$}{Lg}}\label{SSE}
In this subsection we again take $Q=0.5$, but we vary $\alpha$ and $\beta$ independently to determine the effect each coupling has individually. Using the values $Q=0.5$ and $r_h=1$ we find that the bounds for $\alpha$ remain the same as in Subsection \ref{SSC} and $\beta$ can take on any real, positive, value in order to ensure a real initial gradient. We  investigated the case where $\alpha=0.03$ and $\beta$ varies as well as the converse case; we also present the results of varying $\alpha$ and $\beta$ such that $\alpha+\beta$ remains constant, which in this case we took to $0.1$ so that $\alpha$ did not grow too large.

\medskip

\paragraph{Varying \texorpdfstring{$\alpha$}{Lg}}
In this test, we took $\beta=0.03$ and varied $\alpha$ within the range of values that were allowed (that is $\alpha\lesssim0.07$) and above the range of values where there was uncertainty due to the numerical solution. Hence we take the set $\alpha=0.03, 0.05, 0.07$, which gives good indication of the behaviour of the solutions. 

\begin{figure}
    \centering
    \captionsetup{width=0.9\linewidth}
    \includegraphics[width=0.6\linewidth]{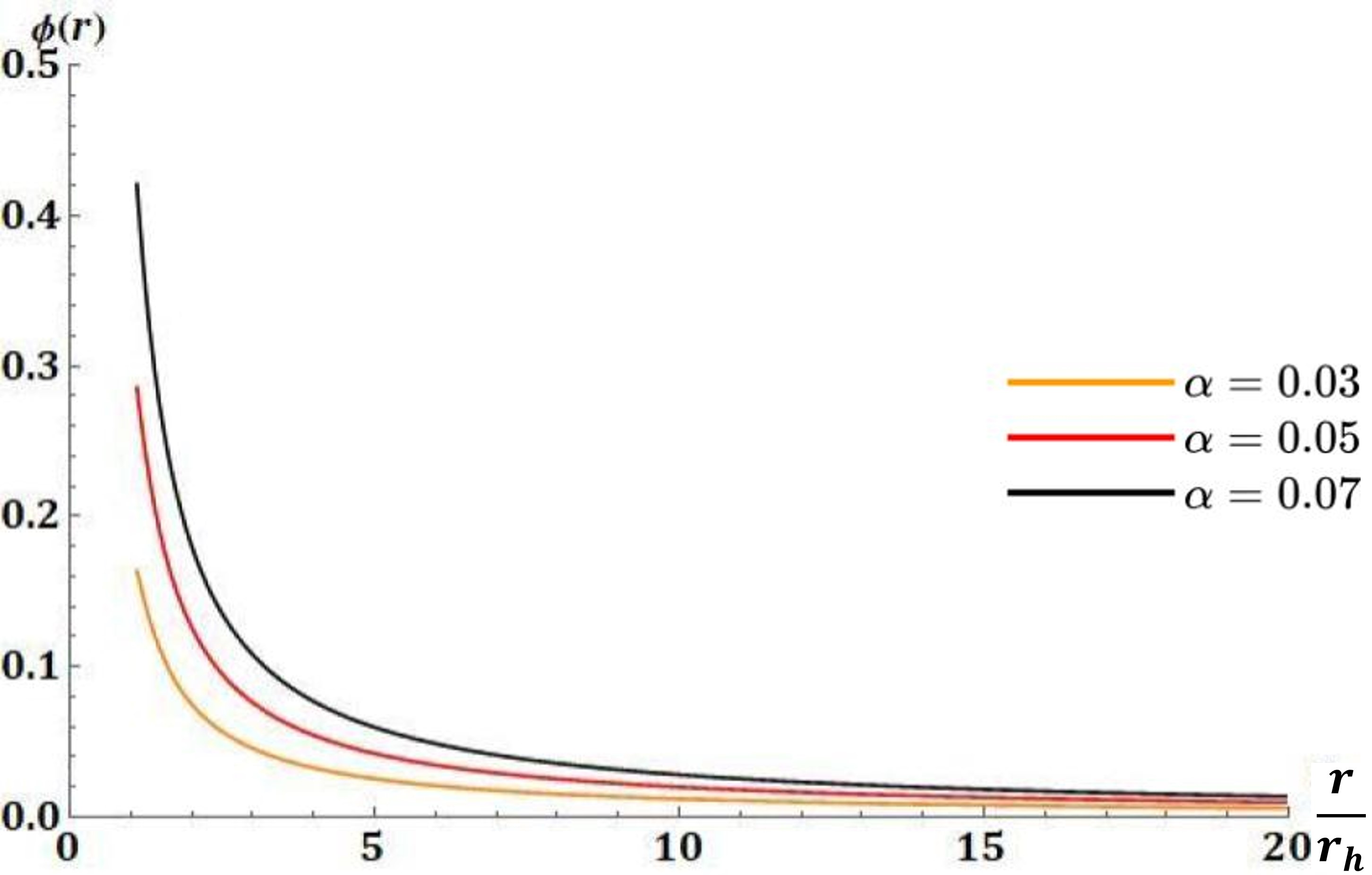}
    \caption{The $\phi$ field plotted for various values $\alpha$ with $\beta=0.03$.}
    \label{fig5}
\end{figure}

The results of these simulations can be found in Figure \ref{fig5}. These results are similar to those in Figure \ref{fig3}, with the exceptions that the very near-horizon values are marginally different, which we can assume is down to the effects of keeping $\beta$ constant and smaller than $\alpha$. The main thing to note here is that we have confirmation that the behaviour described by (\ref{35a}) is accurate, with respect to the description of the scalar charge in terms of $\alpha$. Take for example, the increase of $\alpha$ from 0.03 to 0.05, we would expect the value of the field near the horizon to increase by approximately $1.67$. The numerical solutions tell us that the actual value of field near the horizon increases by a factor of $\sim1.8$, which is approximately in line with \eqref{35a}. Whilst this relation does not hold exactly, it does give us confidence that our assumptions are reasonable.

Furthermore, the results in Figure \ref{fig5} also help us deduce that the value of $\beta$ has a very small effect on the overall scalar field as the difference between the horizon value for $\alpha=\beta=0.05$ and $\alpha=0.05, \beta=0.03$ is very small indeed whilst the increase in the scalar field is relatively large. Thus, we can infer that the coupling to $\mathcal{G}$ has greater scalarisation effects than the gauge field.

\medskip

\begin{figure}
\begin{subfigure}{0.49\textwidth}
    \captionsetup{width=\textwidth}
    \centering
    \includegraphics[width=\textwidth]{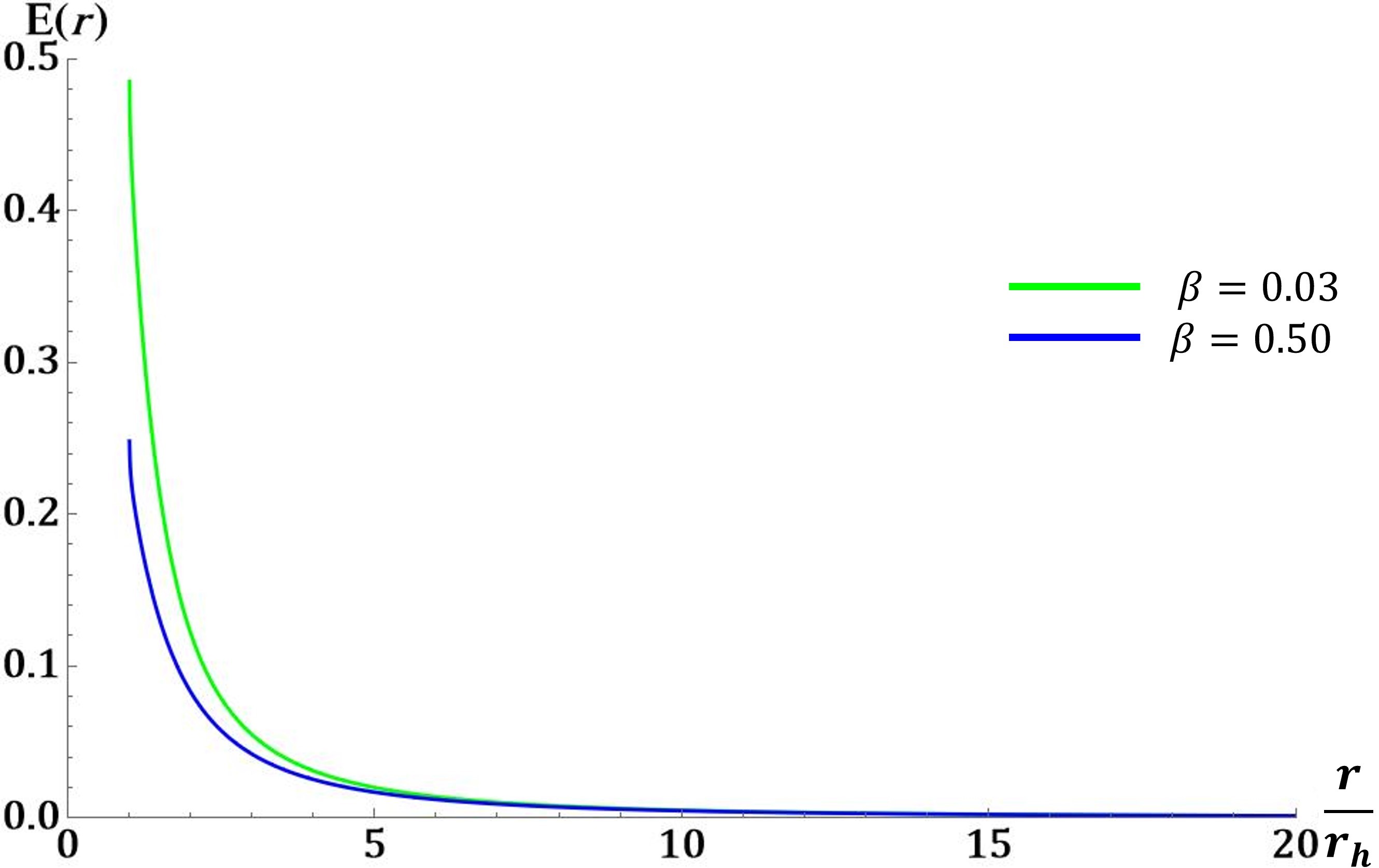}
\caption{The electric field for varying $\beta$ with $\alpha=0.03$.}\label{fig6a}
\end{subfigure}
\hfill
\begin{subfigure}{0.49\textwidth}
    \captionsetup{width=\textwidth}
    \centering
    \includegraphics[width=\textwidth]{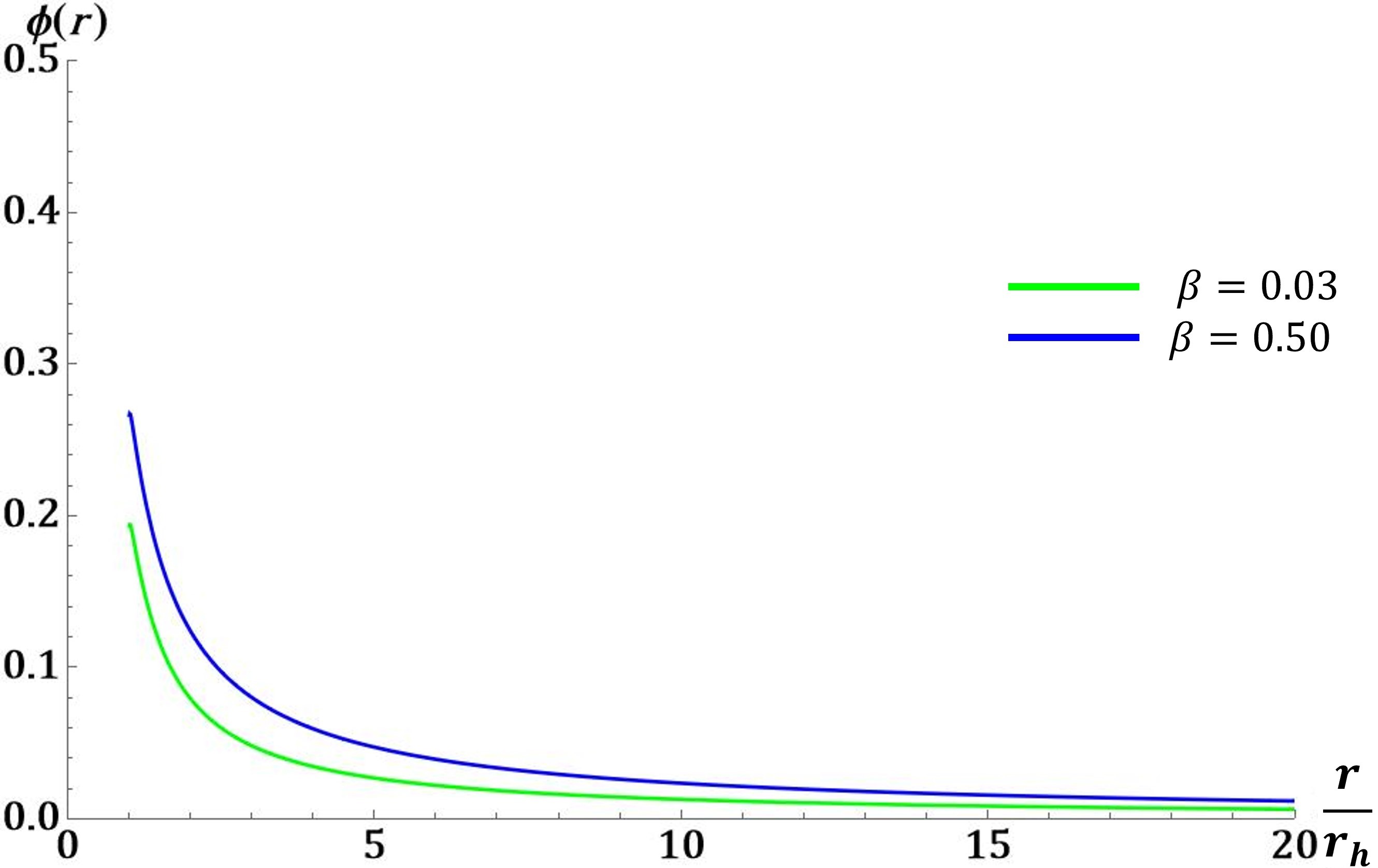}
\caption{The scalar field for various values of $\beta$ with $\alpha=0.03$.}\label{fig6b}
\end{subfigure}
\caption{The results of varying $\beta$ with a constant $\alpha=0.03$.}\label{fig6}
\end{figure}

\paragraph{Varying \texorpdfstring{$\beta$}{Lg}}
We kept $\alpha=0.03$, which is large enough for background effects to be negligible. We investigated $\beta=0.03$ and $0.5$, and the results of this are shown in Figure \ref{fig6}.

Looking at Figure \ref{fig6b} it is evident that the scalar field does not follow, even an approximate, linear relationship in $\beta$ as $\phi$ did in Figure \ref{fig3} for variations of $\alpha=\beta$. This is evident from the form of (\ref{35a}), in which the scalar field's charge depends on the value of $\beta\phi_h$ and since $\phi_h\approx 0.1$, increasing $\beta$ by $\approx16.7$ only leads to an increase of $\approx1.5$ which is in line with our expectations from the perturbative analysis. However, in Figure \ref{fig3} since $\alpha=\beta$, and $\mathcal{O}(\alpha)=\mathcal{O}(\beta)=10^{-2}$, the main contribution arose from the $2\alpha$. This term is always larger than the $8\beta\phi_h Q^2$, term which is of order $10^{-3}$ which is factor of 10 smaller than the $2\alpha$ contribution. This, again, goes some way to confirming the approximation we took in Section \ref{alphaexpansion}. The stronger coupling also leads to a smaller electric field strength, see Figure \ref{fig6a}, although this is unsurprising given the form of \eqref{14}. 

Whilst none of these results are wholly unexpected, it is useful to see that the numerical simulations agree with the theoretical predictions implied by the analysis we performed in Section \ref{S2} and \ref{S3}. It is also reassuring to see that our approximation for $P$ in (\ref{35a}) is again consistent with our numerical results.

\medskip

\begin{figure}[htp]
    \centering
    \captionsetup{width=0.9\linewidth}
    \includegraphics[width=0.6\linewidth]{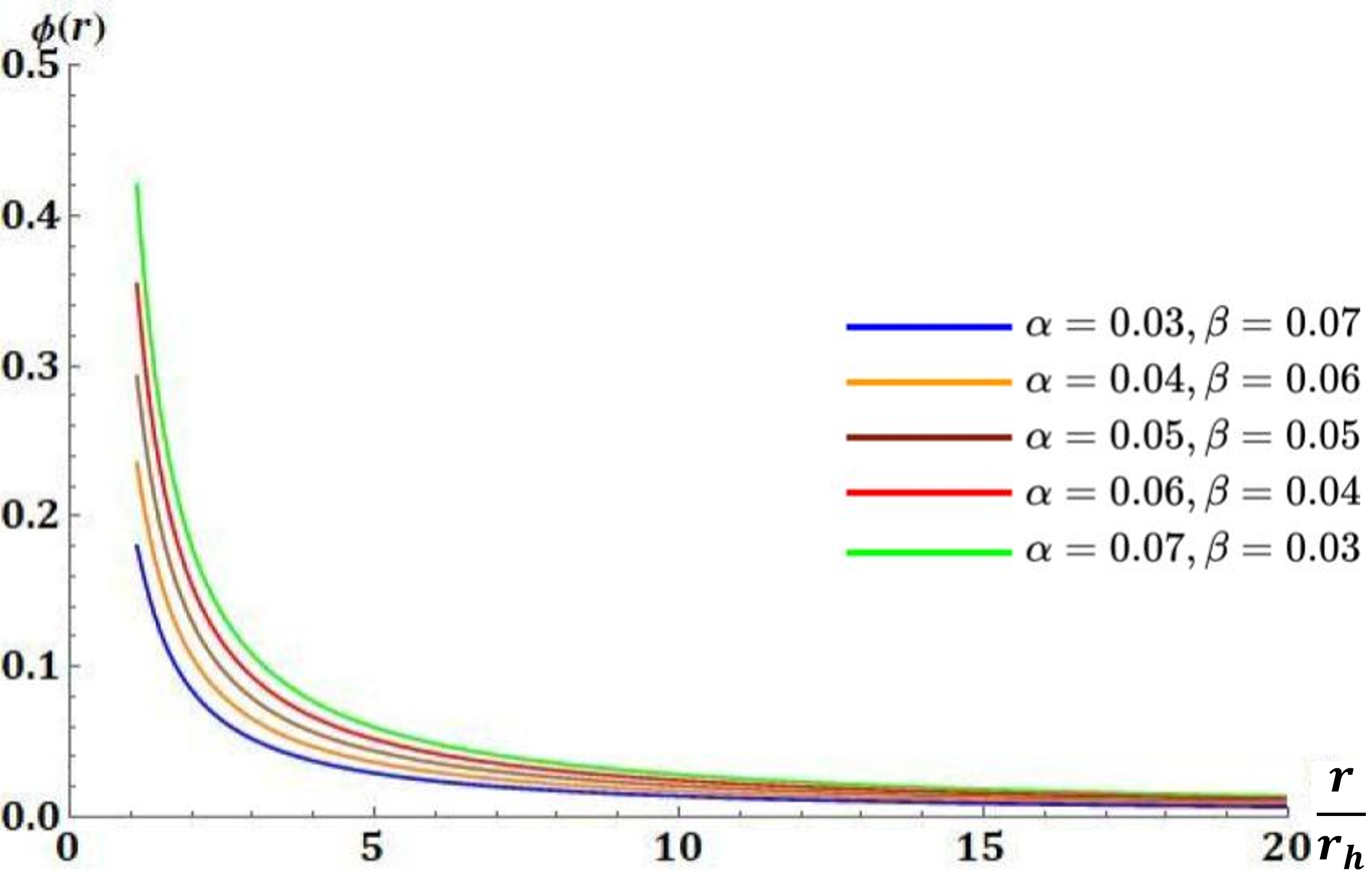}
    \caption{The $\phi$ field plotted for various values $\alpha+\beta=0.1$, as expected larger values of $\alpha$ produce a larger scalar field whereas increasing $\beta$ leads to smaller scalar field due to the form of the scalar charge (\ref{35a}).}
    \label{fig7}
\end{figure}

\paragraph{Varying \texorpdfstring{$\alpha \textit{ and } \beta$}{Lg}}
Here we keep the sum of the couplings equal to a constant such that $\alpha+\beta=0.1$, ensuring that $\alpha$ is not too large such that we can find solutions. The results of this subsection are shown in Figure \ref{fig7}. 

The form of the scalar field confirms the relative importance of the terms in the non-minimally coupled part of the action in (\ref{2}); that is the Gauss-Bonnet coupling has a greater effect on the profile of the scalar field than the gauge field coupling. We can explain this theoretically by invoking (\ref{35a}) again. At the order of the coupling constants and scalar field at the horizon, the $\alpha$ term is approximately 10 times larger than the $\beta$ term. If we compare Figure \ref{fig3} and Figure \ref{fig7} we see that for $\alpha=0.03$, the $\beta=0.07$ coupling only increases the scalar field near the horizon by $\mathcal{O}(10^{-2})$, which is inline with the expected change in the leading order behaviour of the perturbative expansion given by (\ref{35a}). 

\medskip

The main conclusion from the above subsections is that the relative strength of the effects of the Gauss-Bonnet and gauge field terms are an order of magnitude different when the couplings are of the same order of magnitude.
\subsection{Negative \texorpdfstring{$\beta$}{Lg}}\label{negb}
In this subsection we take $\beta<0$ in order to explore the behaviour of the \textit{initial} solutions for negative $\beta$. We have taken $\alpha=0.03$ and $Q=0.5$ in the solutions of this subsection. To work around the issues in this section we manually looked for solutions, for simplicity, initially dropping the asymptotic condition that $\phi \rightarrow 0$. The method employed here was required because for negative $\beta$ there are certain values of the fields for which the differential equations and \eqref{phiprime_h} diverge; $Mathematica$'s $NDSolve$ function looks around the parameter space for the initial conditions and consequently runs into these values which cause issues when trying to solve the equations. To prevent this, we manually explore the parameter space to find the correct solutions - whilst time consuming, it proved the most effective method to tackle this problem.

\begin{figure}
    \centering
    \captionsetup{width=0.9\linewidth}
    \includegraphics[width=0.6\linewidth]{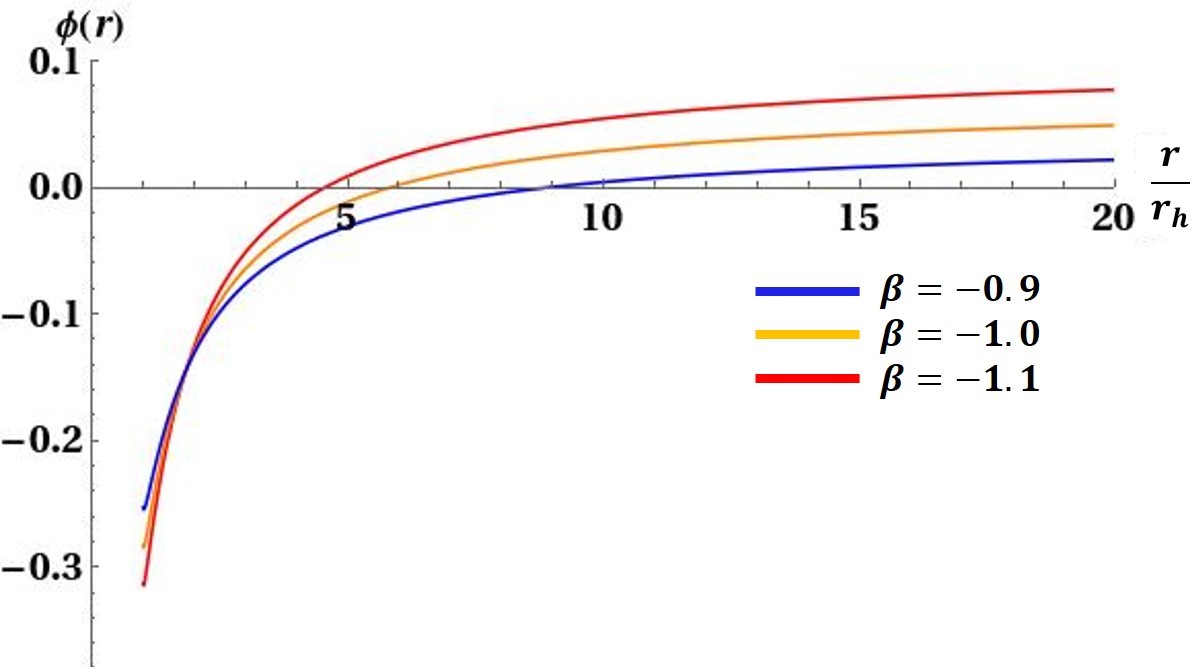}
    \caption{The $\phi$ field plotted for various negative values of $\beta$. Unlike in previous examples the scalar field $\phi$ can take $\phi=0$ at finite range outside of the black hole.}
    \label{fig8}
\end{figure}

The behaviour of $e^A$ and $e^B$ does not change much from the original $\beta>0$ solutions, and they take on the same form as the solutions in previous subsections. The same is true of the electric field, it follows the same structure as laid out in previous figures. However, as one can see in Figure \ref{fig8}, the scalar field does take on unusual behaviour. Firstly, we note that the scalar field can now take a 0 value at a finite distance outside of the black hole, something we had not seen previously. The scalar field also tends to a finite limit, which violates the assumed asymptotic conditions of the theory. However, the limit at large distance is constant and does not diverge. As we shall now show, this allows us to map these solutions to the ones with the correct asymptotic behaviour.
The solutions presented in Figure \ref{fig8} are correct to $\mathcal{O}(0.01)$. This is acceptable since the shooting method in previous calculations had been started from $x=0.01$ away from the black hole.

Solving the equations for negative $\beta$ is quite a difficult task since the solutions are very sensitive to the manual parameter space search. Not only this, but the equations of motion themselves are ill-defined at certain $x$ values for negative $\beta$ and hence there are only certain values $\beta$ can take when less than 0 in order to solve the equations for the range of $x$. It appears physically not very different from the previous profiles, with only a change in sign of the scalar field near the horizon. We can map this solution onto a solution that does satisfy the asymptotic flatness conditions. To do this we shall define new fields, which are related to the old fields, which allow us to enforce the flatness conditions. We shall define,
\begin{align}
    \tilde{F}_{\mu\nu}=qF_{\mu\nu}\label{4.4}\\
    \tilde{\phi}+\phi_\infty=\phi\label{4.5},
\end{align}
where $q$ is a re-scaling of the charge $Q$ and $\phi_\infty$ is the asymptotic value of the scalar field. In this case, we assume to have an action of the form,
\begin{equation}
    \tilde{S}=\int d^4x\sqrt{-g}\bigg(\frac{R}{2}-\frac{1}{8}\tilde{F}_{\mu\nu}\tilde{F}^{\mu\nu}-\frac{1}{2}\partial_\mu\tilde{\phi}\partial^\mu\tilde{\phi}+\tilde{\phi}(\alpha\mathcal{G}-\tilde{\beta}\tilde{F}_{\mu\nu}\tilde{F}^{\mu\nu})\bigg)\label{4.6},
\end{equation}
from which we have to calculate $q$, $\tilde{\beta}$ and $\phi_\infty$, assuming that $\tilde{\phi}\rightarrow0$ as $r\rightarrow\infty$.\footnote{We note that, despite there being a finite scalar field, this is \textit{not} an AdS solution. There are two reasons for this, firstly there is no potential for the scalar field and hence there is no term that can source a cosmological constant. Secondly, the field redefinition only affects the scalar and electric fields, hence the redefintion is not connected to the gravitational part of the theory.} In order to do this we shall compare the actions \eqref{4.6} and \eqref{2} in the asymptotic limit in order to compare the $F_{\mu\nu}F^{\mu\nu}$, noting that we keep the scalar field in the original $\phi$ form of \eqref{2}. This then gives us the scale factor $q$. Then by comparing $\tilde{\beta}\tilde{\phi}\tilde{F}_{\mu\nu}\tilde{F}^{\mu\nu}$ with $\tilde{\phi}\beta F_{\mu\nu}F^{\mu\nu}$ we can obtain the relation between $\beta$ and $\tilde{\beta}$. In doing these comparisons we find that,
\begin{equation}
    \tilde{Q}^2=q^2Q^2, \qquad \tilde{\beta}=\frac{\beta}{q^2}\label{4.7},
\end{equation}
which tells us that the combination $Q^2 \beta$ is invariant and hence defines a unique solution.

\begin{figure}
\begin{subfigure}{0.49\textwidth}
    \captionsetup{width=\textwidth}
    \centering
    \includegraphics[width=\textwidth]{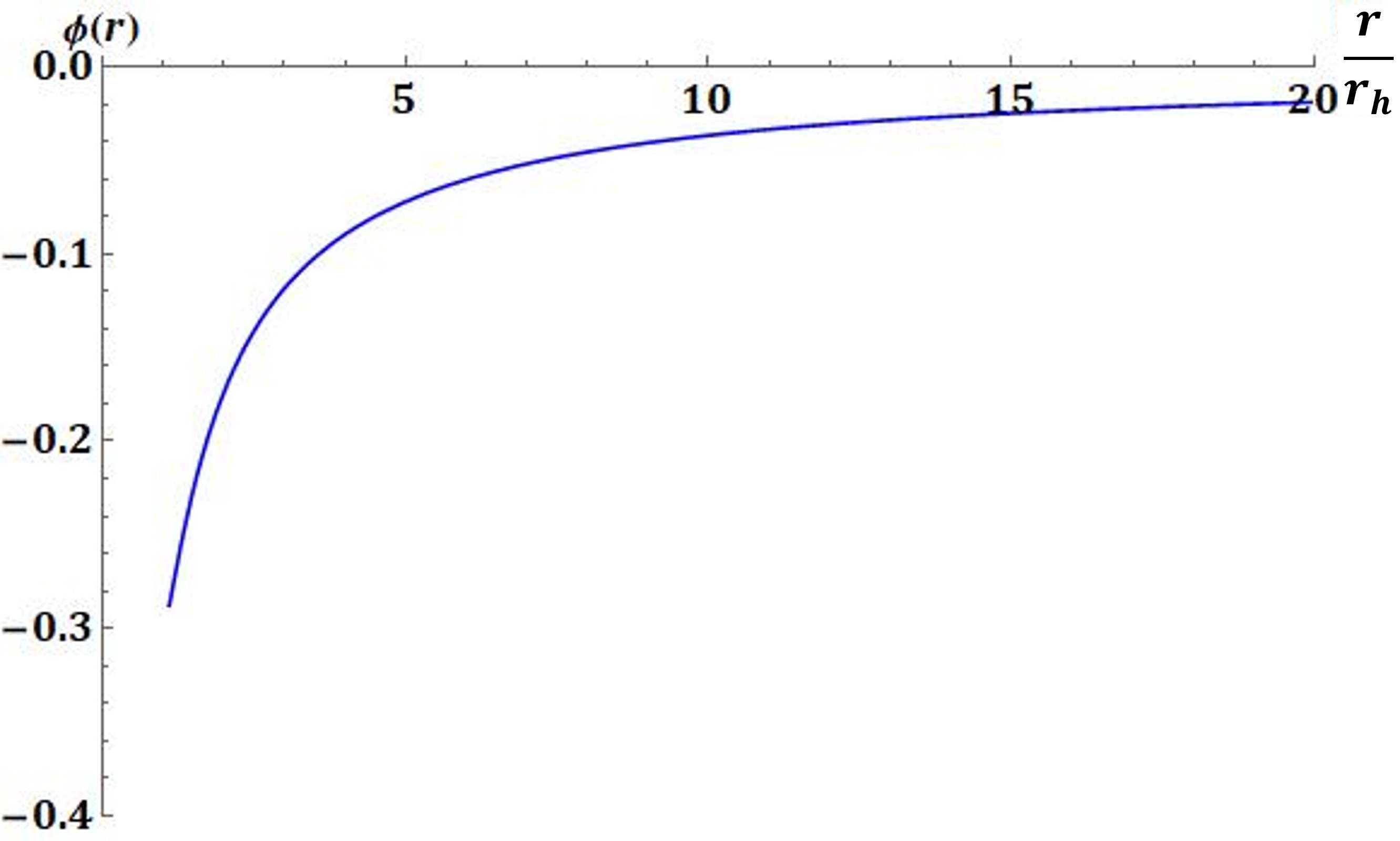}
\caption{The scalar field, after field redefinitions, for $\beta=-0.9$.}\label{fig9a}
\end{subfigure}
\hfill
\begin{subfigure}{0.49\textwidth}
    \captionsetup{width=\textwidth}
    \centering
    \includegraphics[width=\textwidth]{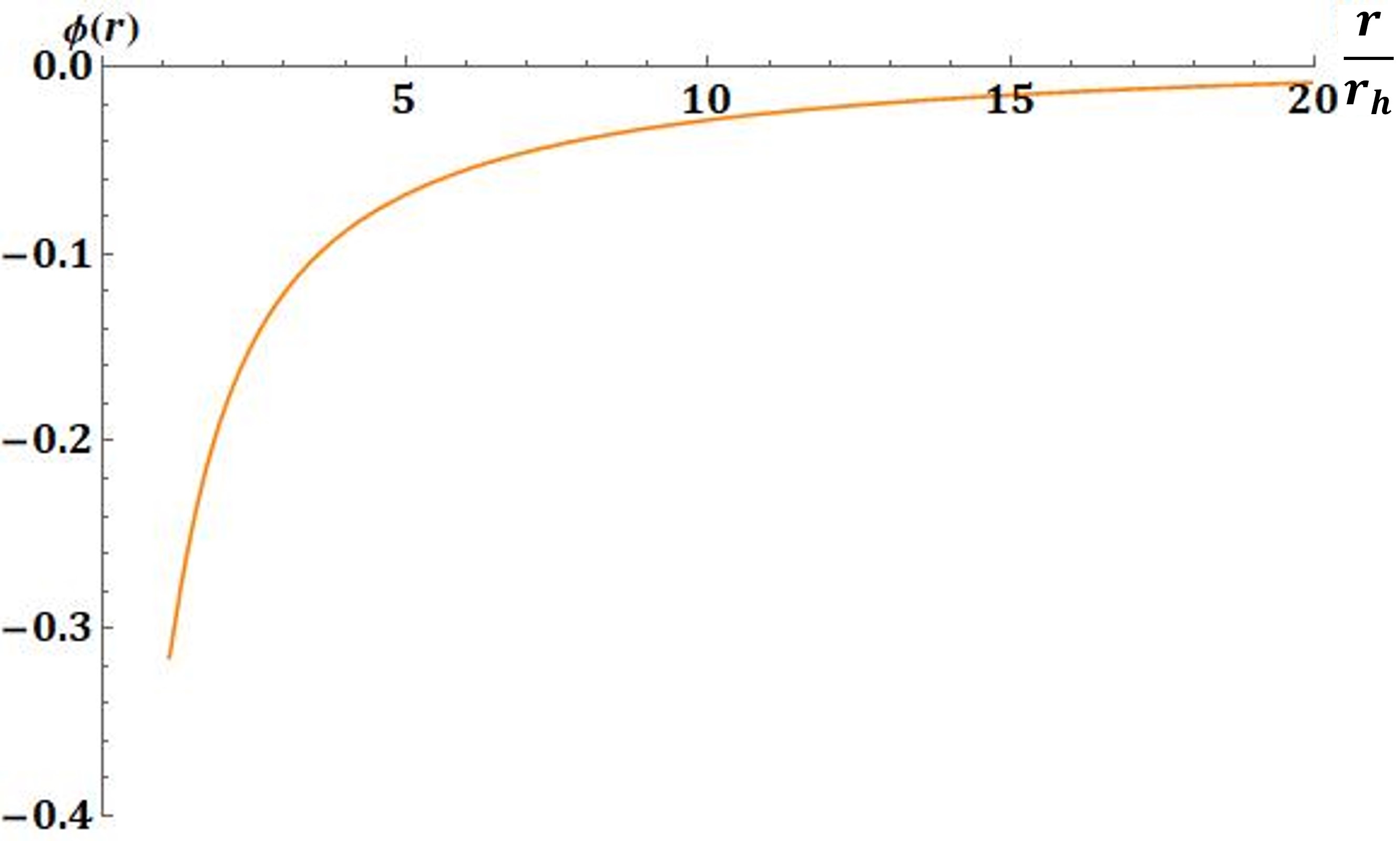}
\caption{The scalar field, after field redefinitions, for $\beta=-1.0$.}\label{fig9b}
\end{subfigure}
\hfill
\begin{center}
\begin{subfigure}{0.49\textwidth}
    \captionsetup{width=\textwidth}
    \centering
    \includegraphics[width=\textwidth]{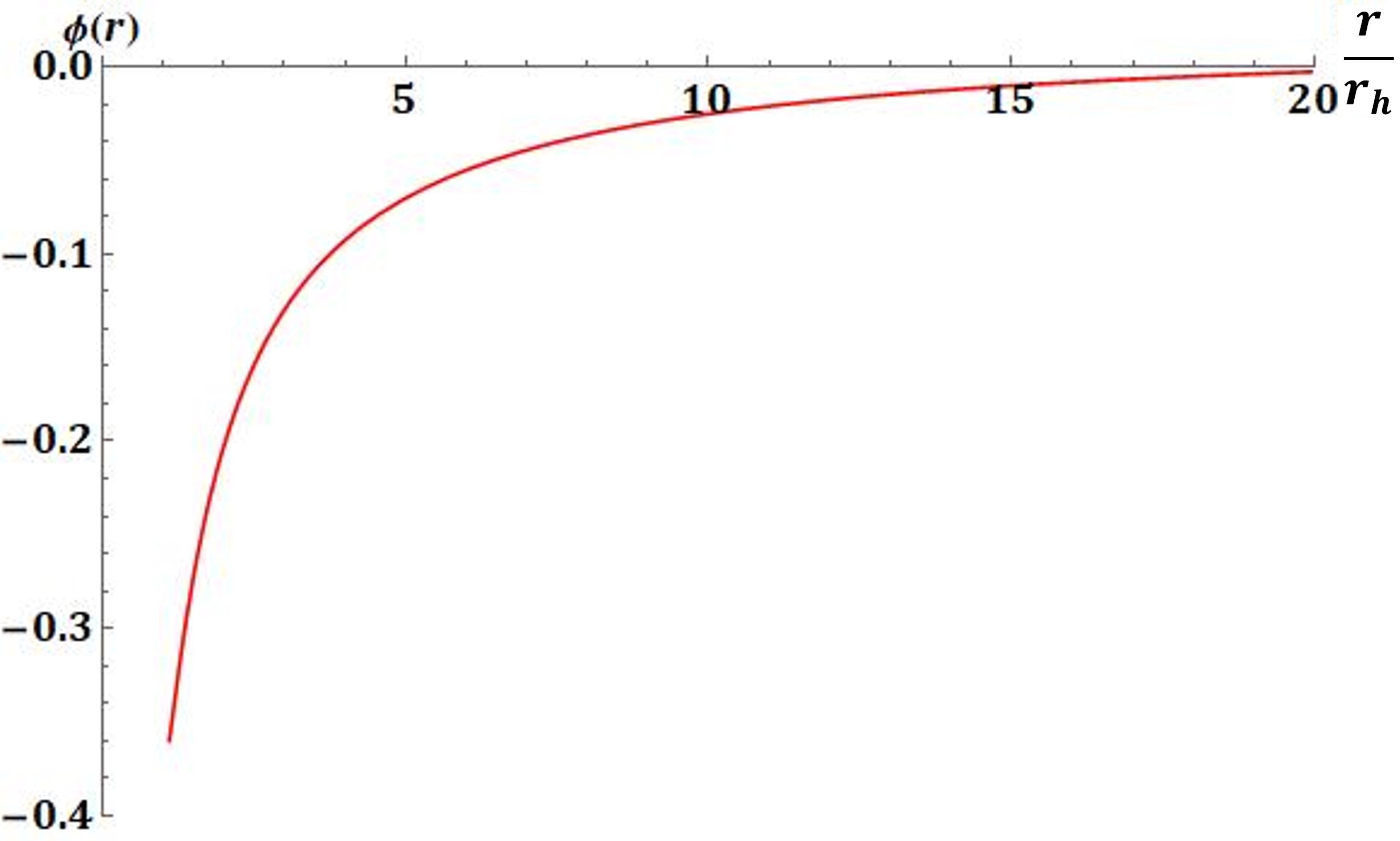}
\caption{The scalar field, after field redefinitions, for $\beta=-1.1$.}\label{fig9c}
\end{subfigure}
\end{center}
\caption{The results of redefining the fields for negative $\beta$. These solutions now obey the boundary conditions of the problem and required the modification of the black hole's charge, $\beta$ coupling and $r_h$.}\label{fig9}
\end{figure}

We focus on the specific case of $\beta=-1$ as an example, which has $\phi\rightarrow0.065$. Then, going through the simple comparison outlined above, we find that $q^2=12/25$. Hence, the new values for $\tilde{\beta}$ and $\tilde{Q}$ are given by,
\begin{equation}
   \tilde{Q}=\pm\sqrt{\frac{3}{25}}, \qquad \tilde{\beta}=-\frac{25}{12},\label{4.8}
\end{equation}
for which we can then solve the equations of motion. We need not discuss the effect on the Gauss-Bonnet term due to the constant shift of $\phi$ since it is topological and so when multiplied by a constant it integrates to 0.

Once we redefine $Q$ and $\beta$ we have to consider how the value of $r_h$ changes. Since the redefinition does not affect the mass of the black hole, the horizon radius will change and this will have to be factored into any numerical solutions that are found. Unfortunately, there are no analytical relations between the values of the parameters owing to the form of the action and hence we can only approximate the change in radius. One way we can make this approximation is to use the Reissner-Nordstr\"om relation from \eqref{31} at $\mathcal{O}(r^{-2})$,
\begin{equation}\label{4.9}
    r_h=\frac{1}{2}(2M\pm\sqrt{4M^2-Q^2}),
\end{equation}
which for $Q=0.5$ and $r_h=1$, gives $M\approx17/32$. Plugging this back in with the new $Q$ value we find that $\tilde{r}_h\approx1.0335$, which is a small deviation from the original value. The solution corresponding to these values of $\tilde{Q}$, $\tilde{\beta}$ and $\tilde{r}_h$ does in fact satisfy the boundary conditions.

We can then repeat the analysis for $\beta=-0.9$ and $-1.1$. Dealing with $\beta=-0.9$ first we find that $\phi\rightarrow0.0247$ as $r\rightarrow\infty$, this gives $q^2=1003/1250$ and hence $\tilde{Q}=0.4479$, $\tilde{\beta}=-1.1216$ and $\tilde{r}_h=1.0130$. For the case of $\beta=-1.1$ we find $\phi\rightarrow0.0877$. This gives $\tilde{Q}=0.2731$, $\tilde{\beta}=-3.686$ and $\tilde{r}_h=1.045$. The scalar field results of all field redefinitions are presented in Figure \ref{fig9}, and it is now easy to see that all solutions obey the asymptotic boundary conditions. We have displayed each solution in its own sub-plot for clarity.

We could of course work to higher orders by solving perturbatively at infinity or near the horizon. We have already done the asymptotic perturbative expansion in \eqref{31} to cubic order. If we used this in place of \eqref{4.9} then we would have to find the new $P$ value and find the roots of the cubic. Hence by redefinition of the fields it is possible to enforce the asymptotic conditions of the problem. This yields new values for $Q$, $\beta$ and $r_h$ which then produce solutions which obey the boundary conditions of the original problem.

\subsection{Shifting the Boundaries}
Previously, we stated that we were able to find solutions by shooting from a point just outside of the horizon, which avoids the divergence in the metric typical of the horizon. For this we shot from $x=0.01$ which corresponds to $1.01 r_h$, however we need to ensure that minimum value of $x$ has minimal impact on the solutions generated by the shooting method. To determine whether the solutions are affected only slightly by the choice of the initial $x$ value we tested $x=\{0.001,0.01,0.1\}$ which correspond to $r=\{1.001,1.01,1.11\}r_h$ respectively. Of course, the smaller the $x$ value, the closer to the horizon we solve and hence the more accurate the solutions we find. To perform these tests we have tested one of the solutions found previously where $Q=1$, $\alpha=\beta=0.05$ and $r_h=1$.

\begin{figure}
    \centering
    \captionsetup{width=0.9\linewidth}
    \includegraphics[width=0.75\linewidth]{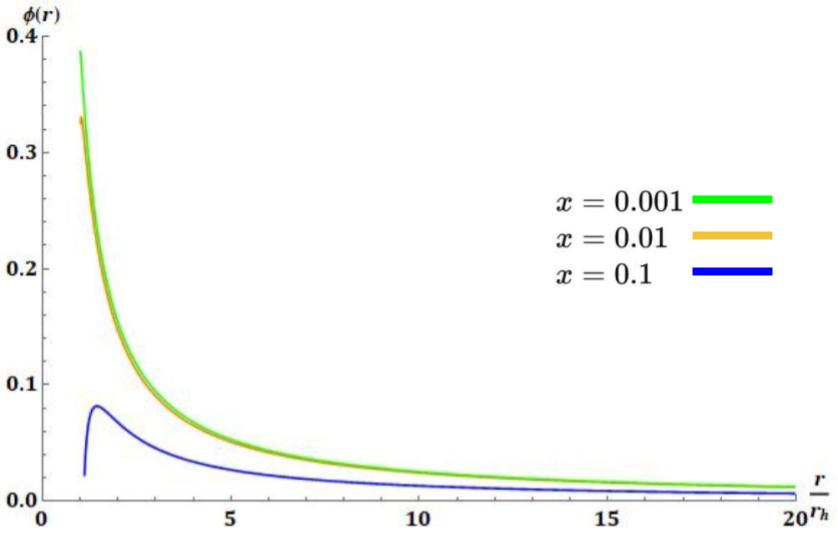}
    \caption{The $\phi$ field plotted for various initial values of $x$. The value of $x=0.1$ shows a dramatic difference from the two smaller values. The two smaller values have very similar profiles but still show a difference very close to the horizon.}
    \label{fig10}
\end{figure}

The results of these tests, for the metric functions and the electric field, were only very slightly changed for each value of $x$, with the $x=\{0.001,0.01\}$ being almost identical (less than $10^{-5}$ difference) and the $x=0.1$ value being different by less than $10^{-3}$. This gives us very good confidence that the $x=0.01$ used in previous results in a good compromise between speed (for larger $x$ values) and accuracy (for smaller $x$) values. A larger difference between $x=\{0.001,0.01\}$ and $x=0.1$ occurs for the scalar field profile, and a minor difference occurs between the two smaller $x$ values. This is not surprising since the initial gradient of the scalar field is determined strongly by the initial $\phi_h$ value which is affected by the $x$ value. The difference between the larger value and the two smaller values for the scalar profile is of $\mathcal{O}(0.1)$ near the horizon and the profile is also quite different. The difference between the two smaller $x$ values is less pronounced and the profile is almost identical with the difference only occurring for the horizon value. This assures us that, for $r\gtrapprox5r_h$, the two solutions for $x=0.001$ and $x=0.01$ are very close. This demonstrates that, below $x\approx0.01$ the profiles of the solutions do not change a great deal. The results for the scalar profile are displayed in Figure \ref{fig10}.

\subsection{Inner Horizons}
Before concluding this section, it is useful to make a comment about the solution within the horizon; theories of the kind explored here have finite size physical singularities that set a limit on the size of the horizon in terms of the parameters of the theory \cite{PhysRevD.90.124063BHhairexample,PhysRevD.54.5049,PhysRevD.65.084014}. It is beyond the scope of this paper to find such solutions within the horizon but it is possible to mention some interesting features of such singularities in this work, for example, \cite{PhysRevD.90.124063BHhairexample,PhysRevD.54.5049,PhysRevD.65.084014} show that the violation of the reality condition for $\phi'_h$ corresponds to the existence of a finite size singularity. It is not possible to derive an exact expression for the radius at which this singularity appears in the theory in this paper as the form of $\chi$ in \eqref{chi_phiprime_h} is too high an order in $r$ to solve. However, as well as being able to easily see this graphically, we can perturbatively solve for $r$. E.g.\ if we assume $\alpha\sim\beta\ll1$ we can take the terms in \eqref{chi_phiprime_h} that are up to order $\alpha^2$, then solving for $r$ we find,
\begin{equation}
    r\gtrapprox \sqrt{2}\sqrt{\frac{2^{11/3}\alpha^2}{(\sqrt{Q^4\alpha^4-1024\alpha^6}-Q^2\alpha^2)^{1/3}}+2^{1/3}(\sqrt{Q^4\alpha^4-1024\alpha^6}-Q^2\alpha^2)^{1/3}}\label{chireality},
\end{equation}
and so in line with the literature, we may find a finite size singularity within the horizon. However, it is possible that certain solutions may not have such finite size singularities since the form of $\phi_h'$ has so many terms it is feasible that all of them make the solution real\footnote{We cannot say anything too precise without a formal numerical solution, but of course the value of $\chi$ in \eqref{chi_phiprime_h} is easily plotted.}. In fact there are many values of $Q$ and $\alpha$, ($Q>4\sqrt{2\alpha}$), which make the above equation reduce to $r>0$, and so there is no \textit{a priori} reason for a finite singularity in this case. Note that \eqref{chireality} is only perturbative and it is the best one can do given the $r-$order of the $\phi'_h$ equation, but we can give an explicit numerical example. Taking one of the first examples where $Q=1$ and $\alpha=\beta=0.05$ it actually transpires that $\phi'_h$ is real for all values of $r$ and so it would suggest that this solution does not have a finite size singularity. In fact plotting $\chi$ for various values of $\alpha,\beta$ and $Q$, as well as the corresponding value of $\phi_h$\footnote{Although the dependency on $\phi_h$'s value is not that big a contributing factor to the zeroes of the function.} one finds that there are many solutions for which $\chi$ is real and there is no obvious evidence of a finite size singularity, although it is also easy to find examples where $\chi$ does become negative.

\section{Conclusions}\label{S5}
In \cite{PhysRevLett.110.241104.scalarhair} a proof was put forward for a no-hair theorem in Galileon-like gravity theories. In this paper, we presented a theory based on the EGBS theory derived in \cite{PhysRevD.90.124063BHhairexample,PhysRevLett.112.251102}, which circumvents the no-hair theorem. Our theory contained an extra, non-minimally coupled, Maxwell term that generated an electric field. We assumed a static, stationary and spherically symmetric metric ansatz. This allowed us to numerically solve the differential equations presented in (\ref{3}-\ref{5}). These numerical solutions confirmed the scalarisation of the black hole.

We initially began by analysing the asymptotic behaviour of the differential equations once we substituted in the ansatz, which gave us an expression for the scalar field gradient at the horizon of the black hole. This was then used as the basis for the shooting method implemented later. We also analysed the asymptotic behaviour as $r\rightarrow\infty$, which informed us of the number of initial parameters that would be needed in order to specify the solution. It also gave an expansion we could use to firstly determine if the parameters we chose gave a black hole that was not overcharged, and then we used the expansion to determine the value of the unknown parameters.

The exact form of the electric field equation \eqref{14} implied the existence of two branches of solutions. The branch we investigated in this paper corresponded to $\beta\phi>-\frac{1}{8}$ and lead to the results in Section \ref{S4}. The other branch was not investigated since the gauge field would take the wrong sign, hence producing unphysical solutions with energy unbounded from below.

We then numerically solved the equations of motion using the shooting method with (\ref{26}) as an initial gradient, transforming into tortoise coordinates such that we could solve over the entire range of $r$. These results were then presented in Section \ref{S4}. The first thing we noted was that the electric charge of the black hole affects all of the fields around the black hole and the spacetime. This is not unexpected since, in the Reissner-Nordstr\"om solution, the charge of the black hole affects the spacetime at order $\frac{1}{r^2}$ and the same occurs here. The spacetime then determines, more so than the electric field, the form of the scalar field. In the case of varying $\alpha$, only the scalar field was affected strongly by the variation. This is down to the linear relationship of $\alpha$ in (\ref{3}), whereas $\alpha$ only appears in the other equation at order $\alpha^{-1}$.

Then we investigated the $\alpha\neq\beta$ case, concluding that the approximation in (\ref{35a}) was approximately satisfied, even near the horizon. We also demonstrated that the Gauss-Bonnet term had a greater effect on the scalar field profile when the couplings are of order $\mathcal{O}(10^{-2})$, which again provides evidence that there is a very direct link between the curvature of spacetime and the profile of the scalar field.

Finally, we took $\beta<0$ and investigated this set of solutions. We concluded that, while these solutions looked different to the previous results, they were physically not all that different from the previous solutions. Finding solutions in this regime is quite difficult and it was numerically simpler to find the solutions presented in Figure \ref{fig8} which do not obey the correct asymptotic boundary condition for the scalar field. However, as we argued, we can then map these to physical solutions (with the correct asymptotics) and the results were presented in Figure \ref{fig9}.

It is interesting to compare the results in this paper to the GR case, the parameters $\alpha$ and $\beta$ define the perturbation from the GR theory and their effect on the metric is to reduce the gradient of the $e^A$ function near the horizon. These parameters have very little effect on the $e^B$ function but this is expected since $e^B$ grows very large near the horizon, even in the GR case, and so small perturbations from the theory will not change this a great deal. As noted in the above, the electric field is only mainly affected by the charge of the black hole and there is very little effect from the non-minimally coupled terms. Of course, the scalar field is directly correlated with the value of $\alpha$ and $\beta$ and these terms, which source the perturbation from GR, source the scalar field. Physical quantities, such as the radius of the innermost stable circular orbit, also only deviate by small amounts for the solutions, hinting that the physical signatures of EMGBS black holes will be very similar to the GR case.

We should note that we have not included rotation in our calculations, but it would be interesting to generalise our results to include rotating black holes since most astrophysical black holes rotate. An investigation of this kind has been carried out for Einstein-Dilaton-Gauss-Bonnet gravity in \cite{PhysRevD.90.044066} and for a massive complex scalar field in \cite{PhysRevLett.112.221101}. It would also be of interest to generalise the coupling $\phi$ to $f(\phi)$, with a different coupling function $g(\phi)$ for the Maxwell field. Finally, it would be interesting to treat the full system as a perturbation from the Schwarzschild metric, as in (\ref{35a}), in order to assess how closely the near horizon behaviour of the system can be modelled by a perturbative expansion in small $\alpha, \beta$.

\section*{Acknowledgements}
CLH would like to thank the Mathematical Sciences department at Durham University for hosting him during the summer of 2019 for the duration of this work. He would also like to thank Prof.\ D.\ J.\ Smith for his continued supervision and guidance throughout this project. This work was supported in part by STFC Consolidated Grant ST/P000371/1.

\appendix

\numberwithin{equation}{subsection}
\renewcommand{\theequation}{A\arabic{equation}}
\section{Divergence properties of \texorpdfstring{$e^{A}$}{Lg} and \texorpdfstring{$A'$}{Lg} near the Horizon}\label{AA}
In this appendix we show that, in the limit of $r\rightarrow r_h$, $e^{-A}\asymp A'$. We do this by showing that the limit,
\begin{equation}\label{A1}
    \lim_{r\rightarrow r_h}\frac{A'}{e^{-A}}=\xi,
\end{equation}
where $\xi$ is some non-zero finite number. In order to do this we must write the near horizon expansion of $e^A$ as a power series in $(r-r_h)$,
\begin{equation}\label{A2}
    \begin{aligned}
    e^A&=a_1(r-r_h)+a_2(r-r_h)^2+...\\
    &=\sum_{n=1}^\infty a_n(r-r_h)^n,
    \end{aligned}
\end{equation}
The derivative of $A$ can then be expressed in terms of (\ref{A2}) and its derivatives as
\begin{equation}\label{A4}
    A'=(e^A)'e^{-A}=a_1e^{-A}+2a_2(r-r_h)e^{-A}+... \; .
\end{equation}
Hence, the limit in (\ref{A1}) is then expressed as
\begin{equation}\label{A5}
    \begin{aligned}
    \lim_{r\rightarrow r_h}\frac{A'}{e^{-A}}&=\lim_{r\rightarrow r_h} \frac{a_1e^{-A}+2a_2(r-r_h)e^{-A}+...}{e^{-A}}\\
    &=a_1,
    \end{aligned}
\end{equation}
which is a finite limit. Hence in the near horizon limit, $e^{-A}$ and $A'$ diverge in very much the same way and we may take the approximation that $e^{-A}\approx A'/a_1$.

\section{Near Horizon Expansion Functions and \texorpdfstring{$\mathcal{O}(\frac{1}{r^4})$}{Lg} approximations}\label{AB}
\renewcommand{\theequation}{B\arabic{equation}}
In this appendix we show the results of the full near horizon expansion, and also the $\mathcal{O}(\frac{1}{r^4})$ expansion coefficients from (\ref{31}-\ref{34}). Whilst these coefficients add little to the discussion, they have been included here for completeness. The order $1/r^4$ term in the asymptotic expansion of $e^A$ is given by,
\begin{equation}\label{B1}
    \begin{aligned}
    &\frac{1}{24r^4}\big[192\alpha MP+M^2P^2-16\beta MPQ^2-P^2Q^2+64\beta^2P^2Q^2+8\beta^2Q^2\big].
    \end{aligned}
\end{equation}
In the case of $e^{B}$ the fourth-order terms take the slightly more complicated form,
\begin{equation}\label{B2}
    \begin{aligned}
    &\frac{1}{48r^4}\big[768M^4-768\alpha MP-416M^2P^2+12P^4-144M^2Q^2+480\beta MPQ^2+14P^2Q^2\\
    &-512\beta^2P^2Q^2+3Q^4-64\beta^2Q^4\big].
    \end{aligned}
\end{equation}
The scalar field has $r^{-4}$ correction of the form,
\begin{equation}\label{B3}
    \begin{aligned}
   &\frac{1}{24r^4}\big[-96\alpha M^2+48M^3P-8MP^3-48\beta M^2Q^2-6MPQ^2+256\beta^2MPQ^2\\
   &+12\beta P^2Q^2-768\beta^3P^2Q^2+3\beta Q^4-64\beta^3Q^4\big].
    \end{aligned}
\end{equation}
The near horizon expansion function $f$ of $\phi''$, defined in (\ref{24}), is given in terms of the electric field $E$ rather than $Q$ and has the rather complicated functional form,
\begin{equation}\label{B4}
    \begin{aligned}
    f=&-\left(4 \alpha  \phi '(r)+r\right)\big[48 \alpha  a_1^2+8 a_1\beta  r^4 E(r)^2 \phi (r) \phi '(r)+a_1 r^4 E(r)^2 \phi '(r)+8 a_1 \beta  r^3 E(r)^2\\
    &+32 \alpha  a_1\beta  r^2 E(r)^2 \phi '(r)-256 \alpha ^2 a_1 \beta  E(r)^2\phi (r) \phi '(r)-32 \alpha ^2 a_1 E(r)^2 \phi '(r)\\
    &+4 a_1^2 r^3 \phi '(r)+16 \alpha  a_1^2 r^2 \phi '(r)^2+16 \beta ^2 r^4 E(r)^4 \phi (r)+2 \beta  r^4 E(r)^4\\
    &-128 \alpha  \beta ^2 r^2 E(r)^4 \phi (r)^2-32 \alpha  \beta  r^2 E(r)^4 \phi (r)-2 \alpha  r^2 E(r)^4\big]\big/a_1\big[-384 \alpha ^2 a_1+4 a_1 r^4\\
    &+16 \alpha  a_1 r^3 \phi '(r)+8 \beta  r^5 E(r)^2 \phi (r)+r^5 E(r)^2-256 \alpha ^2 \beta  r E(r)^2 \phi (r)-32 \alpha ^2 r E(r)^2\big] \; .
    \end{aligned}
\end{equation}
By setting the numerator to 0, we can obtain a finiteness condition on $\phi'$ at the horizon. This produces one of the initial boundary conditions we use to solve the numerical problem. This can be expressed in terms of the charge $Q$ as in (\ref{phiprime_h}). We already have the form of $\chi$, which was given in (\ref{chi_phiprime_h}). Here we give the form of $\psi,\varphi,\zeta$. The form of $\psi$ is,
\begin{equation}\label{B5}
\begin{aligned}
    \psi=& -4 r_h^8 + Q^2 r_h^ 2(32 \alpha^2 + r_h^4) (1 + 8 \beta \phi_h)^{-1}
    + 8\alpha Q^2 (\alpha Q^2 - 8\beta r_h^4) (1 + 8 \beta \phi_h)^{-2}.
    \end{aligned}
\end{equation}
The form of $\varphi$ is given by,
\begin{equation}\label{B6}
    \varphi= 4 r_h^2 - Q^2 (1 + 8 \beta \phi_h)^{-1}.
\end{equation}
Finally, we have the form of $\xi$,
\begin{equation}
    \begin{aligned}
    \zeta=& 32 \alpha r_h^7 - 8\alpha Q^2 r_h(r_h^4 + 32 \alpha^2) (1 + 8 \beta \phi_h)^{-1}
    + 256 \alpha^2 \beta Q^2 r_h^3 (1 + 8 \beta \phi_h)^{-2} .
    \end{aligned}
\end{equation}

\section{Numerical Errors}\label{AppC}
\begin{figure}
\begin{subfigure}{0.49\textwidth}
    \captionsetup{width=\textwidth}
    \centering
    \includegraphics[width=\textwidth]{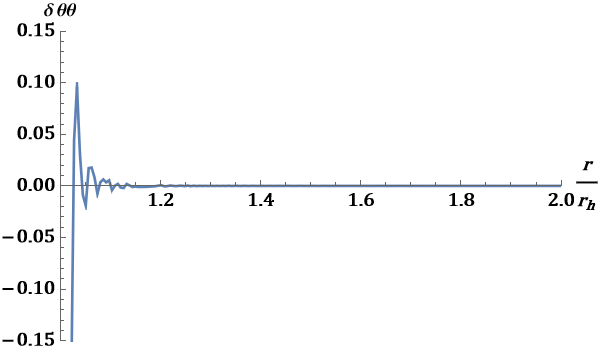}
\caption{The residual of the numerical integration for the $\theta\theta$ Einstein equation.}\label{fig11a}
\end{subfigure}
\hfill
\begin{subfigure}{0.49\textwidth}
    \captionsetup{width=\textwidth}
    \centering
    \includegraphics[width=\textwidth]{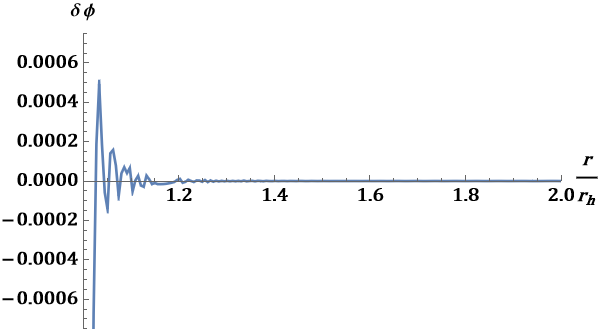}
\caption{The residual of the scalar field equations after numerical integration.}\label{fig11b}
\end{subfigure}
\captionsetup{width=0.9\linewidth}
\caption{The residuals for the two equations used to solve for the solutions in the main body of the paper. Here we take an indicative example in which $\alpha=\beta=0.05$, $Q=1$ and $r_h=1$.}\label{fig11}
\end{figure}

In this appendix, we briefly discuss the numerical errors associated with the integration methods used in the main body of the paper. In order to do this, we take an indicative example and note that all other solutions presented above follow the same sort of pattern of residuals; the indicative example sets $\alpha=\beta=0.05$, $Q=1$ and $r_h=1$. To find the solutions, we solved the $\theta\theta$ and $\phi$ equations of motion once the $E$ field had been eliminated via the solution to its equation of motion; we could have equally solved the $tt$ and $\theta\theta$ equations, or any combination of the three to find the solutions and the variation of the combinations was of the order of the residuals. The residuals of the equations we did solve, $\theta\theta$ and $\phi$, are shown in Figure \ref{fig11}. The residuals are in general quite small everywhere but very close to the horizon, and in the case of the $\phi$ equation the numerical residuals remain less than 0.2\% for the whole range of the solution shown in Figure \ref{fig1b}. It is hard to use the $\theta\theta$ residual to give a percentage as in the $\phi$ case, but it is reasonable to expect that the residual error is of the same order of magnitude. Hence the \textit{numerical errors} are very small and hence the solutions represent accurate results of integration.

\begin{figure}
    \centering
    \captionsetup{width=0.8\linewidth}
    \includegraphics[width=0.75\linewidth]{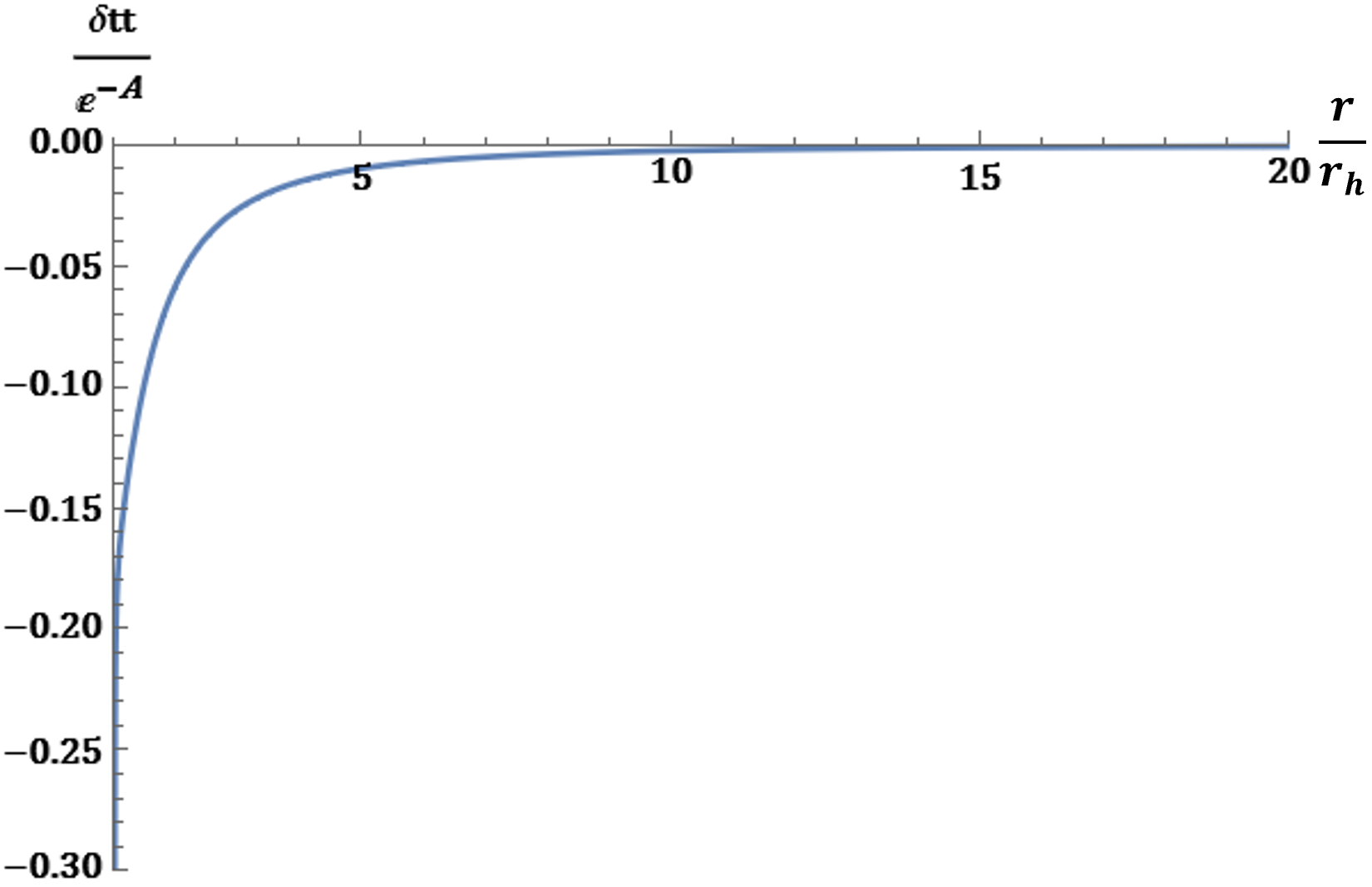}
    \caption{The residual of the $tt$-differential equation divided by $e^{-A}$. This equation was not used in solving for the new solutions.}
    \label{fig_tt}
\end{figure}

That being said, the $tt$ equation that is not used to solve for the solution seems to pose a problem since it begins to diverge to $-\infty$ close to the horizon, however this is to be expected. Satisfying this equation as we approach the horizon requires a precise cancellation between divergent quantities (involving $e^A$ and $B'$) to return a zero result and in light of these large quantities it is not surprising that the residual grows large. However, once normalised by $e^{-A}$ as seen in Figure~\ref{fig_tt} the errors were of order 5\% at around 2$r_h$. 

\section{Virial Identity}
\label{virial}
\renewcommand{\theequation}{D\arabic{equation}}
We can derive a virial identity to perform an additional check on the numerical solutions. Virial
identities are derived from an effective action which is the Lagrangian after substituting in the
ansatz for the solution. In particular, the virial identity arises from the fact that the effective action must be stationary for the classical solution, and specifically considering a variation which rescales the radial coordinate. The method is presented clearly in \cite{Herdeiro:2021teo} where it is also explained that for GR it is necessary to include the Gibbons-Hawking-York boundary term. In our
case we need to generalise this boundary action due to the inclusion of the Gauss-Bonnet term coupled to the scalar $\phi$. The correct boundary action in this case can be found in \cite{Julie:2020vov}.
We then find the virial identity for the action (\ref{2}) with the spherically symmetric ansatz (\ref{7}), (\ref{8}), (\ref{9}) for the metric, scalar field and gauge potential,
\begin{equation}
    \begin{aligned}
\label{VirialIdentity}
    \int_{r_i}^{\infty} dr & \Big( -2 e^{(A+B)/2} + 2 e^{(A-B)/2} - 2 e^{(A-B)/2} B' (r - r_i)
     + 2 \alpha (\phi' + \phi B')\\
     &\times e^{A/2}(e^{B/2} - e^{-B/2}) A'+ \frac{1}{2}r(2r_i - r)(1 + 8\beta\phi)(V')^2 e^{-(A+B)/2}\\
     &+ r(r - 2r_i)(\phi')^2 e^{(A+B)/2}\Big) \\
    = & 
    \left[ 4(r - r_i)e^{A/2}(e^{-B/2} - 1) + 2 \alpha \phi e^{A/2}(e^{B/2} - e^{-B/2}) A' \right]_{r_i}^{\infty}
    \end{aligned}
\end{equation}
where $r_i \ge r_h$ is an arbitrary radius which we take to be $1.5r_h$, as this is far enough from the horizon to not be too sensitive to numerical errors. This identity was tested numerically for $\alpha=\beta=0.05$ and $Q=0.5$ for a number of $r_i$ values. At $r_i=1.1r_h$ we found that the identity was satisfied down to an $\sim2\%$ error\footnote{This is determined as the discrepancy between the left and right sides of the virial identity compared to the magnitude of one side.} indicating that the numerical method is breaking down this close to the horizon, but for $r_i=1.2r_h$ the accuracy increased to give a $\sim1\%$ error and at $r_i=1.5r_h$ the error was of order $\sim0.2\%$. We performed this test on other solutions from the main body of the paper and found that all tested solutions satisfied the above identity to within a discrepancy of order $1\%$ or better for $r_i = 1.2 r_h$ and much less for $r_i = 1.5 r_h$. In fact for smaller values of $\alpha$ and $\beta$ we can get significantly better agreement. E.g.\ for $\alpha = \beta = 0.01$ and $Q = 0.5$ we find the discrepancy is of order $0.1\%$ even for $r_i = 1.1 r_h$. These results give us confidence that the solutions found in the main body of the paper are indeed physical solutions, but consistent with the analysis in appendix~\ref{AppC} we cannot solve too close to the horizon.

Finally, for reference, assuming the spherically symmetric ansatz
(\ref{7}), (\ref{8}), (\ref{9}) for the metric, scalar field and gauge potential for an
asymptotically flat solution with action,
\begin{equation}
\label{GenAction}
    \begin{aligned}
    S=&\frac{M_P^2}{16\pi}\int d^4x \sqrt{-g}\bigg( R - \frac{1}{4}h(\phi)F_{\mu\nu}F^{\mu\nu} - \partial_\mu\phi\partial^\mu\phi + f(\phi)\mathcal{G} \bigg),
    \end{aligned}
\end{equation}
the virial identity is,
\begin{equation}
    \begin{aligned}
\label{GenVirialIdentity}
    \int_{r_i}^{\infty} dr & \Big( -2 e^{(A+B)/2} + 2 e^{(A-B)/2} - 2 e^{(A-B)/2} B' (r - r_i)
     + (f'(\phi)\phi' + f(\phi) B')e^{A/2}\\
    & \times(e^{B/2} - e^{-B/2}) A'+ \tfrac{1}{2}r(2r_i - r)h(\phi)(V')^2 e^{-(A+B)/2}
     + r(r - 2r_i)(\phi')^2 e^{(A+B)/2} \Big)\\
    = & 
    \left[ 4(r - r_i)e^{A/2}(e^{-B/2} - 1) + f(\phi) e^{A/2}(e^{B/2} - e^{-B/2}) A' \right]_{r_i}^{\infty} \; .
    \end{aligned}
\end{equation}
Here $f(\phi)$ and $h(\phi)$ are arbitrary functions while in the specific theory considered
in this paper we chose $f(\phi) = 2\alpha \phi$ and $h(\phi) = 1 + 8 \beta \phi$.

\newpage

\bibliography{bibliography}

\end{document}